\documentclass[twocolumn,aps,prc,tightenlines,floats,floatfix,nofootinbib,preprintnumbers]{revtex4}

\def\st#1{{\kern-1pt} \not\!#1}

\def\sp{\kern +3pt}
\def\sm{\kern -3pt}

\def\spQ{\kern +6pt}

\def\bea{\begin{eqnarray}}
\def\eea{\end{eqnarray}}

\def\sfrac#1#2{{\textstyle \frac{#1}{#2}}}

\newcommand{\bra}[1]{\langle #1|}
\newcommand{\ket}[1]{|#1\rangle}

\def\be{\begin{equation}}
\def\ee{\end{equation}}
\def\ba{\begin{eqnarray}}
\def\ea{\end{eqnarray}}

\usepackage{graphics}
\usepackage{graphicx}
\usepackage{epsf}
\usepackage{amsmath}
\usepackage{amssymb}

\begin{document}

\phantom{0}
\vspace{-0.2in}
\hspace{5.5in}

\preprint{ADP-11-44/T766, YITP-11-104, J-PARC-TH-003}

\vspace{-1in}

\title
{\bf Valence quark and meson cloud
contributions for
the $\gamma^\ast \Lambda \to \Lambda^\ast$
and $\gamma^\ast \Sigma^0 \to \Lambda^\ast$
reactions}
\author{
G. Ramalho$^1$,  D. Jido$^{2,3}$, and  K. Tsushima$^4$
\vspace{-0.1in}  }

\affiliation{
$^1$CFTP, Instituto Superior T\'ecnico,
Universidade T\'ecnica de Lisboa,
Av. Rovisco Pais, 1049-001 Lisboa, Portugal
\vspace{-0.15in}
}

\affiliation{
$^2$Yukawa Institute for Theoretical Physics,
Kyoto University, Kyoto 606-8502, Japan
\vspace{-0.15in}
}

\affiliation{
$^{3}$J-PARC Branch, KEK Theory Center, Institute of Particle and Nuclear Studies,
High Energy Accelerator Research Organization (KEK), 203-1,
Shirakata, Tokai, Ibaraki, 319-1106, Japan
\vspace{-0.15in}
}

\affiliation{
$^4$CSSM, School of Chemistry and Physics,
University of Adelaide, Adelaide SA 5005, Australia
}

\vspace{0.2in}
\date{\today}

\phantom{0}

\begin{abstract}
We estimate the valence quark contributions for the
$\gamma^\ast Y \to \Lambda^\ast$ ($Y=\Lambda,\Sigma^0$) electromagnetic
transition form factors.
We focus particularly on the case $\Lambda^\ast = \Lambda(1670)$
as an analog reaction with $\gamma^\ast N \to N(1535)$.
The results are compared with those obtained from
chiral unitary model, where the $\Lambda^\ast$
resonance is dynamically generated
and thus the electromagnetic structure comes directly
from the meson cloud excitation of the baryon ground states.
The form factors for the case $Y=\Sigma^0$ in particular, depend crucially
on the two real phase (sign) combination, a phase between the $\Lambda$ and $\Lambda^\ast$
states, and the other, the phase between the $\Lambda$ and $\Sigma^0$ radial wave functions.
Depending on the combination of these two phases, the form factors
for the $\gamma^\ast \Sigma^0 \to \Lambda^\ast$ reaction can be enhanced or suppressed.
Therefore, there is a possibility to determine the phase combination by experiments.
\end{abstract}

\vspace*{0.9in}  
\maketitle

\section{Introduction}

With the development of the modern accelerators
the study of the meson and light baryon structure
has been becoming one of the most exciting topics in
physics.
Although the underlying theory of strong interaction,
quantum chromodynamincs (QCD),
is known for a long period, its complexity in low energy region
forces us to use some effective theories
(aside from lattice QCD),
either based on the quark and gluon degrees of freedom,
or some effective interactions between
the mesons and baryons.
Among the various possible meson-baryon reactions,
the reactions that involve strangeness are particularly
interesting, due to the accessibility of the modern
accelerators to strange particles such as kaons, $K$,
antikaons, $\bar{K}$, and hyperons,
$\Sigma$ and $\Lambda$.

In this study we focus on the
electromagnetic excitations of $\Lambda$ hyperon
ground state. The $\Lambda$ ground state, $\Lambda(1116)$
is $J^P = \sfrac{1}{2}^+$ and belongs to the spin 1/2
octet baryon multiplet, in which also the nucleons belong.
The lowest mass of the $\Lambda$ excited state ($\Lambda^\ast$)
reported by the particle data group~\cite{PDG}
is $\Lambda(1405)$, a $J^P= \sfrac{1}{2}^-$ state.
The $\Lambda(1405)$ state
has collected a lot of interests over the years
for the reasons following:
(i) it has been suggested as a dynamically generated state
(molecular-like state) composed largely of
the $\pi \Sigma$ and $\bar K N$
states~\cite{Dalitz60,Dalitz67,Wyld67,Kaiser95};
(ii) it is difficult to classify in terms of
naive quark models based on $SU(6)$ symmetry.
In the representation of spin-flavor $SU(6)$ symmetry
the $\Lambda(1405)$ state can be a mixture
of three distinct 3-quark states
including the $\Lambda$-singlet state~\cite{Isgur78,Darewych83,Capstick00}.
However, its mass is difficult to predict
in Karl-Isgur model~\cite{Capstick00},
as well as in cloudy bag model (CBM)~\cite{Veit85}.
In CBM the $\Lambda(1405)$ was interpreted primarily
as a $\bar K N$ bound state~\cite{Veit85}.
Thus, there is a strong indication that
the $\Lambda(1405)$ is a dynamically generated
meson-baryon molecular-like state with a single or
a double pole structure~\cite{Veit85,Oset:1997it,Jido03,Hyodo:2007jq,Hyodo:2008xr,Hyodo11,Ikeda11,Doring11,Doring10,Oset:2001cn,Oller:2000fj}.
In particular, it was demonstrated
that the $\Lambda(1405)$ is composed substantially of the
meson-baryon components within the chiral unitary model~\cite{Hyodo:2008xr}.
Nevertheless, there are some works that support
the $\Lambda(1405)$ as a 3-quark state~\cite{Leinweber90,An10,Menadue11}.

Therefore, to study the $\gamma^\ast \Lambda \to \Lambda^\ast$ reaction
is very interesting also by the reasons following.
In one aspect this reaction has a possible
analogy with the $\gamma^\ast N \to N^\ast(1535)$ reaction.
Because $\gamma^\ast \Lambda \to \Lambda^\ast$ is a transition
between a $J^P=\sfrac{1}{2}^+$ and a $J^P=\sfrac{1}{2}^-$ states,
we have a possibility to interpret the $\Lambda(1405)$ as a
$p$-state excitation of one quark in
the ground state $\Lambda(1116)$,
analogous to $N^\ast(1535)$,
a $p$-state excitation of the nucleon~\cite{S11}.
However, the $\Lambda(1405)$
has considerably lower mass than $N^\ast(1535)$.
Furthermore, it has a larger
mass difference with the nearest $d$-state partner $\Lambda(1520)$
compared to the case of $N^\ast(1535)$ and $N^\ast(1520)$.
The mass order is even reversed for the $\Lambda(1405)$ case.
Because of the reasons discussed above,
it is very difficult to interpret naively $\Lambda(1405)$
as a simple $p$-state excitation of $\Lambda(1116)$.

Searching for the next higher mass excited state of $\Lambda(1116)$
with $J^P = \sfrac{1}{2}^-$, one finds $\Lambda(1670)$,
which can be an analogous with $S_{11}$ excitation
of the nucleon, $N^\ast(1535)$.
Since $\Sigma^0$ is the neutral $\Sigma$ ground state
($J^P=\sfrac{1}{2}^+$) which belongs to the spin 1/2
octet baryon multiplet, and the $\gamma^\ast \Sigma^0 \to \Lambda^\ast$
reaction is similar to the $\gamma^\ast \Lambda \to \Lambda^\ast$ reaction,
we also focus on the $\gamma^\ast \Sigma^0 \to \Lambda^\ast$ reaction
in this study.
Because the $\Lambda(1116)$ and $\Sigma^0(1193)$ are
similar in masses, the two reactions differ
mainly in the initial state quark configurations.
As for the other interesting aspect, we note that the
$\Lambda(1670)$ resonance can also be described as a dynamically
generated meson-baryon state~\cite{Oset:2001cn,Oller:2000fj},
and the $\gamma^\ast Y \to \Lambda^\ast$ 
transition form factors for $Y=\Lambda,\Sigma^0$, were calculated
in chiral unitary model~\cite{Doring10}.

In the previous works, a valence quark model
was applied to study the $\gamma^\ast N \to N^\ast(1535)$ reaction,
and the corresponding transition form
factors and helicity amplitudes were
studied~\cite{S11,S11scaling}.
The reaction was also studied in
a coupled-channels chiral dynamics (chiral unitary model)~\cite{Jido08}.
In the chiral unitary model the contributions
for the transition form factors come entirely
from the meson-baryon states (meson cloud effect).
For the $\gamma^\ast N \to N^\ast(1535)$
reaction the transition form factors $F_1^\ast$
(Dirac-type) and $F_2^\ast$ (Pauli-type)
can be expressed in terms of the transverse ($A_{1/2}$)
and longitudinal ($S_{1/2}$) helicity
amplitudes~\cite{S11,Aznauryan11a}.
In Ref.~\cite{S11}, it was found that the $F_1^\ast$ can be explained
very well just taking into account the valence quark effect.
By contrast, the meson cloud seems to play
a very important role for the $F_2^\ast$,
in particular in the low $Q^2$ region~\cite{Jido08}.
Then, such different roles between the valence
quark and meson cloud effects may be reflected in the
experimentally extracted helicity amplitudes
$S_{1/2}$ and $A_{1/2}$.
This possibility was indeed demonstrated in Ref.~\cite{S11scaling}.
We will briefly review also these results.

Therefore, one of our main motivations of this study is
to investigate whether or not the different roles of the valence quark
and meson cloud effects observed for the $\gamma^\ast N \to N^\ast(1535)$ reaction,
can also be observed in the
$\gamma^\ast Y \to \Lambda^\ast$ reactions with $Y=\Lambda$ and $\Sigma^0$.
In particular, we focus on the structure of $\Lambda(1670)$ in this study.
Assuming that $\Lambda(1670)$ is a radial $p$-state excitation
of $\Lambda(1116)$, we estimate the valence quark contributions
for the $\gamma^\ast Y \to \Lambda^\ast$ 
transition form factors as well as the helicity amplitudes.
For this purpose, we use the covariant spectator
quark model~\cite{S11,Nucleon,OctetEMFF,ExclusiveR},
which was successfully applied to the study of
the $\gamma^\ast N \to N^\ast(1535)$ reaction.
The results of the covariant spectator quark model
for the $\gamma^\ast Y \to \Lambda^\ast$ reaction
are also compared with those obtained with the
chiral unitary model~\cite{Doring10},
where the $\Lambda^\ast$ is generated as a
meson-baryon molecular-like state
such as the $N \bar{K}$, $\Lambda \eta$ and $\Xi K$ states.
Then, one of the interests is the structure of the $\Lambda(1670)$ state,
namely, how it can be interpreted, either it is
predominantly a meson-baryon molecular-like
state, or dominated by the 3-valence-quark state.
Furthermore, we also show that the
$\gamma^\ast \Sigma^0 \to \Lambda^\ast$
transition form factors depend crucially
on the combination of the two unknown real phases (signs),
a phase between the $\Lambda$ and $\Lambda^\ast$
three-quark wave functions (to be denoted by $\eta_{\Lambda^\ast}$),
and a phase between the $\Lambda$ and $\Sigma^0$ wave functions
(to be denoted by $\eta_{\Lambda \Sigma^0}$).

This article is organized as follows.
In Sec.~\ref{secDefinitions}
we define the $\gamma^\ast Y \to \Lambda^\ast$ ($Y=\Lambda,\Sigma^0$)
transition form factors, and their relations
with the helicity amplitudes.
In Sec.~\ref{secSQM} we present
the covariant spectator quark model
and estimate the valence quark contributions
for $\gamma^\ast Y \to \Lambda^\ast$ ($Y=\Lambda,\Sigma^0$).
We discuss in Sec.~\ref{secCUM} the $\Lambda(1670)$
state based on the chiral unitary model, and estimate
the contributions from the meson-baryon states
in the $\gamma^\ast Y \to \Lambda^\ast$ ($Y=\Lambda,\Sigma^0$) reactions.
In Sec.~\ref{secResults} we present the
results from the both models, and give discussions.
Finally in Sec.~\ref{secConclusions}
we give conclusions of the present study.

\section{Form factors and helicity amplitudes}
\label{secDefinitions}

The $\gamma^\ast Y \to \Lambda^\ast$
electromagnetic transition current
for $Y$ a strangeness $S=-1$
and $J^P=\sfrac{1}{2}^+$ state,
and $\Lambda^\ast$ a $J^P= \sfrac{1}{2}^-$ excited state of
the $\Lambda$ ground state ($J^P=\sfrac{1}{2}^+$),
can be represented as~\cite{S11,Aznauryan11a}
\be
J_Y^\mu= e
\left[
\left(\gamma^\mu -\frac{\not \! q q^\mu}{q^2}\right) F_1^Y(Q^2)
+ \frac{i \sigma^{\mu \nu} q_\nu}{M_{\Lambda^\ast}+ M_Y} F_2^Y(Q^2)
\right]
\gamma_5,
\label{eqJS}
\ee
where $F_i^Y$ ($i=1,2$) are the transition form factors,
and $q$ the four-momentum transfer (defined below) with $Q^2 = -q^2$.
The factor $e$ is the absolute electron charge given by
$e=\sqrt{4\pi \alpha}$
with $\alpha$ being the electromagnetic fine structure constant. 
Note that the form factors are frame independent
since Eq.~\eqref{eqJS} is Lorentz covariant.
We are particularly interested in the cases
$Y=\Lambda$ and $\Sigma^0$ in this study.

The current $J_Y^\mu$ can be projected
on the initial state $u_Y(P_-,S_z)$
and final state $\bar u_{\Lambda^\ast}(P_+, S_z^\prime)$
Dirac spinors, where $P_-$ ($P_+$) is the initial (final)
momentum, \mbox{$q = P_+ - P_-$}, and $S_z$ ($S_z^\prime$)
the spin projection.

More familiar matrix elements
may be the helicity amplitudes.
In this case the current $J_Y^\mu$
is projected on the photon polarization states $\epsilon_\mu^{(\lambda)}$,
where the polarizations can be longitudinal ($\lambda=0$)
or transverse ($\lambda= \pm$).
As the photon polarizations depend on the frame
the helicity amplitudes are frame dependent.
The most common choice of the reference frame
is the final state rest frame, $\Lambda^\ast$ at rest.
In this frame we can define the transverse ($A_{1/2}^Y$)
and longitudinal ($S_{1/2}^Y$) helicity amplitudes as~\cite{Aznauryan11a}
\ba
& &
\hspace{-.5cm}
A_{1/2}^Y= \sqrt{\frac{2\pi \alpha}{K}} \frac{1}{e}
\left< \Lambda^\ast,S_z^\prime=+ \frac{1}{2} \right|
\epsilon^{(+)} \cdot J_Y
\left| Y, S_z= - \frac{1}{2} \right>, \nonumber \\
& & \label{eqA120} \\
& &
\hspace{-.5cm}
S_{1/2}^Y= \sqrt{\frac{2\pi \alpha}{K}} \frac{1}{e}
\left< \Lambda^\ast,S_z^\prime=+ \frac{1}{2} \right|
\epsilon^{(0)} \cdot J_Y
\left| Y, S_z= + \frac{1}{2}  \right>
\frac{|{\bf q}|_Y}{Q},
\nonumber \\
& &
\label{eqS120}
\ea
with $\alpha = \sfrac{e^2}{4\pi}$, and
\be
K= \frac{M_{\Lambda^\ast}^2-M_Y^2}{2M_{\Lambda^\ast}}.
\ee
In the above $|{\bf q}|_Y$ is the absolute value of
the photon three-momentum ${\bf q}$
in the $\Lambda^\ast$ rest frame,
\ba
|{\bf q}|_Y=
\frac{\sqrt{[(M_{\Lambda^\ast}+M_Y)^2+Q^2][(M_{\Lambda^\ast}-M_Y)^2+Q^2]
 }{}}{2M_{\Lambda^\ast}}.
\label{eqqY}
\ea
The subindex $Y$ is to label the initial state.

In the $\Lambda^\ast$ rest frame
we can relate the helicity amplitudes
with the form factors~\cite{Aznauryan11a}:
\ba
\hspace{-1.2cm}
& &
A_{1/2}^Y=
-2b
\left[F_1^Y + \frac{M_{\Lambda^\ast}-M_Y}{M_{\Lambda^\ast}+M_Y} F_2^Y
\right],
\label{eqA12X}
\\
\hspace{-1.2cm}
& &
S_{1/2}^Y=
\sqrt{2}b(M_{\Lambda^\ast}+M_Y) \frac{|{\bf q}|_Y}{Q^2}
\left[\frac{M_{\Lambda^\ast}-M_Y}{M_{\Lambda^\ast}+M_Y} F_1^Y - \tau F_2^Y
\right],
\label{eqS12X}
\ea
with
\be
\tau= \frac{Q^2}{(M_{\Lambda^\ast}+M_Y)^2}, \label{eqtau}
\ee
and
\be
b= e \sqrt{\frac{(M_{\Lambda^\ast}+M_Y)^2+ Q^2}{8M_Y (M_{\Lambda^\ast}^2-M_Y^2)}}.
\ee

\section{Spectator quark model}
\label{secSQM}

In the spectator formalism~\cite{Gross,Stadler98,Gross11b}
a baryon is represented as a 3-quark state
and the wave function is expressed in terms of the
corresponding baryon spin-flavor~\cite{Nucleon,Omega,ExclusiveR}.
Then the baryon system is decomposed as
an off-mass-shell constituent quark, free to interact with
electromagnetic fields, and two on-mass-shell quarks.
Integrating over the on-mass-shell momenta
we reduce the baryon to a quark-diquark system
which has an effective diquark mass $m_D$~\cite{Nucleon,Omega,Gross11a}.

The electromagnetic interaction with the quark
is described in terms of a vector
meson dominance (VMD) parametrization
(to be described later), that simulates the
constituent quark internal structure.
The quark structure parameterizes
effectively the interactions with gluons
and quark-antiquark polarization effects.
The quark electromagnetic current was calibrated previously
by the nucleon and decuplet baryon data~\cite{Nucleon,Omega},
and also tested in the lattice regime for the nucleon elastic reaction
as well as for the $\gamma N \to \Delta$
transition~\cite{OctetEMFF,Omega,Lattice,LatticeD}.
The model was also applied to the physical regime
to study the octet baryon and decuplet baryon
systems~\cite{OctetEMFF,OctetMU,NDelta,NDeltaD,GE2Omega,DeltaDFF,DeltaDFF2},
and some of the excited states of nucleon
and $\Delta$~\cite{S11,Roper,Delta1600}.

\subsection{Wave functions}

We next discuss the
spin-flavor-radial wave functions
of the systems relevant in this work, namely,
$\Lambda$, $\Sigma^0$, and $\Lambda^\ast$,
where $\Lambda^\ast$ is interpreted as a 3-quark excitation
of the $\Lambda$ ground state with negative parity.
The structure of the $\Lambda$ and $\Sigma^0$
systems are based on Ref.~\cite{OctetEMFF},
which studied the octet baryon electromagnetic properties.
As for the $\Lambda^\ast$, based on the structure
considered for the $N^\ast(1535)$ in Ref.~\cite{S11},
we generalized it.

\begin{table*}[t]
\begin{center}
\begin{tabular}{l c c c}
\hline
\hline
$Y$   & $\ket{M_S}_Y$  & &  $\ket{M_A}_Y$  \\
\hline
$\Lambda$ &
$\sfrac{1}{2}
\left[ (dsu-usd) + s (du-ud)
\right]$
& &
$\sfrac{1}{\sqrt{12}}
\left[
s (du-ud) - (dsu-usd) -2(du-ud)s
\right]$ \\
$\Sigma^0$ &
$\sfrac{1}{\sqrt{12}}
\left[
s (du+ud) +(dsu+usd) -2(ud+du)s
\right]$
& &
$\sfrac{1}{2}
\left[ (dsu+usd) -s (ud+du)
\right]$ \\
\hline
\hline
\end{tabular}
\end{center}
\caption{Flavor wave functions of $\Lambda$ and $\Sigma^0$.}
\label{tablePHI}
\end{table*}

\subsubsection{Spin-flavor wave functions}

In Ref.~\cite{OctetEMFF} it was shown that
the octet baryon systems can be described reasonably well
in an S-state configuration for the quark-diquark system,
and that the same structure of wave function applies
for all the members of the octet baryons except for the flavor states.
The general structure of the wave function is written by~\cite{OctetEMFF}
\ba
\Psi_Y(P,k) =
\frac{1}{\sqrt{2}} \left[
\phi_S^0  \ket{M_A}_Y + \phi_S^1  \ket{M_S}_Y
\right] \psi_Y(P,k),
\label{eqPsiL}
\ea
where $P$ is the total momentum of particle $Y$,
$k$ the diquark momentum,
$\psi_Y$ is the radial wave function, and
$\phi_S^{0,1}$ are the spin wave functions.
The flavor wave functions are presented in Table~\ref{tablePHI}.
The spin wave functions (the same for all the octet members)
are expressed as~\cite{OctetEMFF,Nucleon,S11},
\ba
& &
\phi_S^0= u_Y(P,S_z), \nonumber \\
& &
\phi_S^1= - (\varepsilon_P^{\ast})_\alpha (\lambda_D)U_Y^\alpha(P,S_z),
\label{SpinWF}
\ea
where $U_Y^\alpha$ is the vector-spinor~\cite{Nucleon,NDelta},
\ba
U_Y^\alpha(P,s)=
\frac{1}{\sqrt{3}} \gamma_5
\left(
\gamma^\alpha - \frac{P^\alpha}{M_Y}
\right) u_Y(P,S_z),
\label{eqUa}
\ea
and $u_Y(P,S_z)$ the $Y$-Dirac spinor
with the spin projection $S_z$.
In Eq.~(\ref{SpinWF}), $\varepsilon_P(\lambda_D)$
with $\lambda_D=0,\pm$ are the
\mbox{spin-1} diquark polarization states
defined in the fixed-axis representation
as a function of the $Y$-momentum~\cite{Nucleon,FixedAxis}.
For later discussions, we note that,
even the flavor states can be well defined,
the total wave functions can have sign ambiguities
due to the normalization constants for the radial wave functions $\psi_Y$.
Note however, that the sign is not relevant for the elastic
reactions like the $\Lambda$ and $\Sigma$  electromagnetic
form factors, since the results are proportional
to the integral with the product of the two (real number)
functions $\psi_Y$, although they
have different arguments~\cite{OctetEMFF}.

As for the $\Lambda^\ast$ state with $J^P= \sfrac{1}{2}^-$,
we use the analogy with the $N^\ast(1535)$
to represent the corresponding wave function.
Assuming as in Ref.~\cite{S11} that
the $\Lambda^\ast$ is dominated by the internal
quark states with a total spin 1/2
and has no $P$-states inside the diquark
(pointlike diquark), it is written by
\ba
\Psi_{\Lambda^\ast} (P,k) =
\frac{1}{\sqrt{2}} \left[
\Phi_\rho  \ket{M_A}_\Lambda - \Phi_\lambda  \ket{M_S}_\Lambda
\right] \psi_{\Lambda^\ast} (P,k),
\label{eqPsiLS}
\ea
where $\Phi_\rho$ and $\Phi_\lambda$
are the spin states to be defined shortly,
and $\psi_{\Lambda^\ast}$ the $\Lambda^\ast$
radial wave function.
Note that the $\Lambda^\ast$ and $\Lambda$ are described
by the same flavor wave function (see Table~\ref{tablePHI}).
The states $\Phi_{\rho,\lambda}$ are defined
respectively by~\cite{S11},
\ba
\hspace{-.5cm}
& &
\Phi_\rho(\pm)
= -\gamma_5 {\cal N}_{\Lambda^\ast}^\prime
\left[ ( \varepsilon_0 \cdot \tilde k ) u_{\Lambda^\ast} (\pm)
- \sqrt{2} (\varepsilon_\pm \cdot \tilde k )  u_{\Lambda^\ast} (\mp)
\right], \nonumber \\
\hspace{-.5cm}
& &
\Phi_\lambda(\pm)
= \gamma_5 {\cal N}_{\Lambda^\ast}^\prime
\left[ ( \varepsilon_0 \cdot \tilde k ) \varepsilon_\alpha^\ast
U_{\Lambda^\ast}^\alpha (\pm)
- \sqrt{2} (\varepsilon_\pm \cdot \tilde k )
 \varepsilon_\alpha^\ast
U_{\Lambda^\ast}^\alpha (\mp)
\right], \nonumber \\
\hspace{-1cm}
& &
\label{eqPHI}
\ea
where $\pm$ hold for the spin projections $S_z= \pm \sfrac{1}{2}$,
and $\varepsilon_0$ is a short notation for $\varepsilon_P(0)$ of
the diquark polarization
associated with the $\Lambda^\ast$, and
${\cal N}_{\Lambda^\ast}^\prime$ is the normalization factor, and
\be
\tilde k= k -\frac{P \cdot k}{M_{\Lambda^\ast}^2} P.
\ee
This last four-momentum reduces to the diquark
three-momentum in the $\Lambda^\ast$ rest frame.
As for $U_{\Lambda^\ast}^\alpha$, it is defined
by Eq.~(\ref{eqUa}) with $M_Y \to M_{\Lambda^\ast}$.
The normalization factor ${\cal N}_{\Lambda^\ast}^\prime$
can be represented as
\ba
{\cal N}_{\Lambda^\ast}^\prime= \eta_{\Lambda^\ast} N,
\label{eqNorma}
\ea
where $N= 1/\sqrt{-\tilde k^2}$,
and $\eta_{\Lambda^\ast}$ is a relative phase (sign)
between the $\Lambda$ and $\Lambda^\ast$ states
to be discussed later.

The states $\Phi_\rho$ and $\Phi_\lambda$ in Eq.~(\ref{eqPHI})
are constructed respectively
to be antisymmetric and symmetric for the
interchange of the quarks 1 and 2~\cite{S11}.

\subsubsection{Radial wave functions}

Since the baryon and the diquark are on-mass-shell
in the spectator quark model, we can represent
the baryon radial wave function in term of $(P-k)^2$.
We can use then the dimensionless variable,
\ba
\chi_B= \frac{(M_B-m_D)^2-(P-k)^2}{M_B m_D},
\label{eqCHI}
\ea
where $M_B$ is the mass of the baryon $B$
and $m_D$ the diquark mass.
Following Ref.~\cite{OctetEMFF},
we take the form for the $Y$ ($=\Lambda, \Sigma^0$) wave functions,
\ba
\psi_Y(P,k) =
\frac{{\cal N}_Y}{m_D(\beta_1 + \chi_Y) (\beta_3 + \chi_Y)},
\label{eqPsiR}
\ea
where the values of $\beta_1$ and $\beta_3$
were fixed in Ref.~\cite{OctetEMFF}
as $\beta_1= 0.0440$ and $\beta_3= 0.7634$, and
${\cal N}_Y$ is the normalization constant.
We assume that ${\cal N}_Y$ is {\it positive}.
While $\beta_1$ parameterizes
the spacial long-range distribution of the quarks
which are dominated by the light quarks,
$\beta_3$ regulates the short-range structure
in a system with only one strange quark.
The normalization constant, ${\cal N}_Y$, is determined
by the condition,
\ba
\int_k |\psi_Y( \bar P, k)|^2 = 1,
\label{eqPsiYN}
\ea
where $\bar P=(M_Y,0,0,0)$ is the $Y$-momentum
in its rest frame, and $\int_k$ stands for
$\int \frac{d^3 {\bf k}} {2E_D (2\pi)^3}$,
where $E_D$ is the diquark on-mass-shell energy.
Note that Eq.~(\ref{eqPsiYN}) only determines
the magnitude of ${\cal N}_Y$, but not the sign. 

For the $\Lambda^\ast$ wave function, we take also the
form of Eq.~(\ref{eqPsiR}), except that $\chi_Y$
is replaced by $\chi_{\Lambda^\ast}$, meaning
that the $M_Y$ is replaced by $M_{\Lambda^\ast}$.
This choice is equivalent to state
that the $\Lambda$ and $\Lambda^\ast$
have the same radial wave function, 
and they are distinguished only by the spin states.
It also means that the normalization
constants in the radial wave functions,
${\cal N}_\Lambda$  and ${\cal N}_{\Lambda^\ast}$
are equal\footnote{For
the radial wave functions with
the structure of Eqs.~(\ref{eqCHI}) and~(\ref{eqPsiR}),
the normalization condition given by Eq.~(\ref{eqPsiYN})
uses $\chi_B= 2\left(\frac{E_D}{m_D}-1 \right)$,
which is independent of the baryon mass.
As a consequence, the normalization
constant is independent of the baryon mass.}
and therefore have the same sign.

We can now discuss the sign $\eta_{\Lambda^\ast}$
in Eq.~(\ref{eqNorma}).
Since it is already assumed that the normalization constant
of the radial wave function ${\cal N}_\Lambda$ is positive,
$\eta_{\Lambda^\ast}$ defines the sign of the
$\gamma^\ast \Lambda \to \Lambda^\ast$ transition form factors.
Because we have no clue for the
sign of the form factors till the date,
we will keep the factor $\eta_{\Lambda^\ast}$
in the following equations.
We call also attention that the sign corresponding to
the $\gamma^\ast N \to N^\ast(1535)$ reaction
is equivalent to $\eta_{\Lambda^\ast}=1$,
where this was determined by the experimentally extracted
sign for the form factor $F_1^\ast$~\cite{S11}.

\subsection{Electromagnetic transition current}

The electromagnetic current for the transition
$\gamma^\ast Y \to \Lambda^\ast$
in a relativistic impulse
approximation is given by~\cite{Nucleon,OctetEMFF,Omega}
\ba
J_Y^\mu=
3 \, e \sum_{\Gamma} \int_k \bar \Psi_{\Lambda^\ast}(P_+,k) j_q^\mu \Psi_Y(P_-,k),
\label{eqJ1}
\ea
where $\Gamma = \left\{ s,\lambda_D \right\}$
(the scalar diquark $s$, and the vector
diquark polarizations $\lambda_D=0,\pm 1$),
and $j_q^\mu$ is the quark current operator associated
with the quark 3.
The factor 3 accounts for the contributions
from the quark pairs (13) and (23)
[the same contribution as that from the pair (12)].

\subsubsection{Quark current}

The quark current $j_q^\mu$ (in $e$ units)
can be represented 
as~\cite{Nucleon,NDelta,Omega,OctetEMFF,OctetMU}
\ba
j_q^\mu = j_1 \hat \gamma^\mu
+ j_2 \frac{i \sigma^{\mu \nu} q_\nu}{2M},
\label{eqJq}
\ea
where $M$ is the nucleon mass, and
\ba
\hat \gamma^\mu = \gamma^\mu - \frac{\not\! q q^\mu}{q^2},
\ea
and $j_i$ ($i=1,2$)
are the Dirac and Pauli quark operators, respectively.
The inclusion of the term  $\not\!\! q q^\mu / q^2$
in the quark current is equivalent
to use the Landau prescription~\cite{Kelly98,Batiz98}
to the final electromagnetic current.
The term restores current conservation but
does not affect the observables calculated~\cite{Kelly98}.

The operators $j_i$ ($i=1,2$) act on the flavor states
$\ket{M_A}_Y$ and $\ket{M_S}_Y$ written in terms
of the symmetry with respect to the quark 3.
The operators $j_i$ can be decomposed
into the sum of $SU(3)$-space operators~\cite{Omega,OctetEMFF},
\ba
j_i = \frac{1}{6}
f_{i+} \lambda_0 + \frac{1}{2} f_{i-} \lambda_3
+ \frac{1}{6} f_{i0} \lambda_s,
\label{eqJi}
\ea
where $\lambda_0= \mbox{diag}(1,1,0)$, $\lambda_3=\mbox{diag}(1,-1,0)$
and $\lambda_s= \mbox{diag}(0,0,-2)$,
and $f_{i\,n}$ ($i=1,2$, $n=0,\pm$)
define the constituent quark form factors.
The operators act on the third quark, where the quark
wave function is represented by $q=(uds)^T$.

The quark electromagnetic form factors
are normalized as
$f_{1\,n}(0)=1$ ($n=0,\pm$), $f_{2\pm}(0)=\kappa_\pm$
and $f_{20}(0)=\kappa_s$.
The isoscalar ($\kappa_+$) and isovector ($\kappa_-$)
anomalous magnetic moments are related
with the $u$ and $d$ quark anomalous magnetic moments
by $\kappa_+= 2 \kappa_u - \kappa_d$
and $\kappa_-= \sfrac{2}{3} \kappa_u - \frac{1}{3}\kappa_d$ \cite{Nucleon}.
As for $\kappa_s$ it is the strange quark anomalous
magnetic moment~\cite{Omega,GE2Omega}.

\subsubsection{Quark electromagnetic form factors}

To parameterize the quark current (\ref{eqJi}),
we adopt the structure inspired by the vector meson
dominance (VMD) mechanism as in Refs.~\cite{Nucleon,Omega}:
\ba
& &
\hspace{-1cm}
f_{1 \pm} = \lambda_q
+ (1-\lambda_q)
\frac{m_v^2}{m_v^2+Q^2} + c_\pm \frac{M_h^2 Q^2}{(M_h^2+Q^2)^2},
\nonumber \\
& &
\hspace{-1cm}
f_{1 0} = \lambda_q
+ (1-\lambda_q)
\frac{m_\phi^2}{m_\phi^2+Q^2} + c_0 \frac{M_h^2 Q^2}{(M_h^2+Q^2)^2},
\nonumber \\
& &
\hspace{-1cm}
f_{2 \pm} = \kappa_\pm
\left\{
d_\pm  \frac{m_v^2}{m_v^2+Q^2} + (1-d_\pm)
\frac{M_h^2 }{M_h^2+Q^2} \right\}, \nonumber \\
& &
\hspace{-1cm}
f_{2 0} = \kappa_s
\left\{
d_0  \frac{m_\phi^2}{m_\phi^2+Q^2} + (1-d_0)
\frac{M_h^2}{M_h^2+Q^2}  \right\},
\label{eqQff}
\ea
where $m_v,m_\phi$ and $M_h$ are the masses respectively
corresponding to the light vector meson $m_v \simeq m_\rho$,
the $\phi$ meson (associated with an $s \bar s$ state),
and an effective heavy meson
with mass $M_h= 2 M$ to represent
the short-range phenomenology.
For the isoscalar component it should be
$m_v=m_\omega$, but we neglect the small mass difference
between the $\rho$ and $\omega$ mesons, and use $m_\rho$.
The coefficients $c_0,c_\pm$ and $d_0,d_\pm$ were
determined in the previous studies of the nucleon
(model II)~\cite{Nucleon} and $\Omega^-$~\cite{Omega}.
The values are respectively,
$c_+= 4.160$, $c_-= 1.160$, $d_+=d_-=-0.686$,
$c_0=4.427$ and $d_0=-1.860$~\cite{Omega}.
The parameter $\lambda_q=1.21$ is fixed to give
the correct quark number density in deep inelastic
scattering~\cite{Nucleon}.

In this study we use the values of the parameters
determined by the study of the octet baryon
electromagnetic form factors~\cite{OctetEMFF}:
\ba
\kappa_u= 1.6690, \hspace{.2cm} \kappa_d= 1.9287, \hspace{.2cm}
\kappa_s= 1.4620.
\ea

With the wave functions (\ref{eqPsiL})~and~(\ref{eqPsiLS})
one can write the quantity in Eq.~(\ref{eqJ1}) as
\ba
\sum_{\Gamma} \bar \Psi_{\Lambda^\ast} j_q^\mu \Psi_Y &=  & +
\frac{{\cal A}}{2}
\left\{
j_1^A \bar \Phi_\rho \hat \gamma^\mu \phi_S^0
+ j_2^A \bar \Phi_\rho  \frac{i \sigma^{\mu \nu} q_\nu}{2M} \phi_S^0
\right\} \nonumber \\
& &
-\frac{{\cal A}}{2}
\left\{
j_1^S \bar \Phi_\rho \hat \gamma^\mu \phi_S^1
+ j_2^S \bar \Phi_\rho  \frac{i \sigma^{\mu \nu} q_\nu}{2M} \phi_S^1
\right\}, \nonumber \\
& &
\label{eqSum}
\ea
where
\ba
& &
j_i^A= {_{\Lambda}\!}\bra{M_A}j_i \ket{M_A}_Y, \nonumber \\
& &
j_i^S=  {_{\Lambda}\!}\bra{M_S}j_i \ket{M_S}_Y,
\ea
for $i=1,2$, and they are the coefficients
that encapsulate the flavor effect~\cite{OctetEMFF,OctetMU,Omega},
and ${\cal A}=
{\cal N}_{\Lambda^\ast}^\prime \psi_{\Lambda^\ast}(P_+,k)   \psi_Y(P_-,k)$.
In Eq.~(\ref{eqSum})
the sum in the diquark polarizations $\lambda_D$ is implicit
for the vector diquark contributions
(terms in $\phi_S^1$).

The calculation of the coefficients
$j_i^{A,S}$ ($i=1,2$) gives,
\ba
j_i^S= \frac{1}{6} f_{i+}, \hspace{.7cm}
j_i^A= \frac{1}{18} (f_{i+} - 4 f_{i0}),
\label{eqJAS}
\ea
for the $\gamma^\ast \Lambda \to \Lambda^\ast$ reaction, and
\ba
j_i^S= - \frac{1}{\sqrt{12}} f_{i-}, \hspace{.7cm}
j_i^A=  \frac{1}{\sqrt{12}} f_{i-},
\label{eqJAS2}
\ea
for the $\gamma^\ast \Sigma^0 \to \Lambda^\ast$ reaction.
In the case of $\gamma^\ast \Lambda \to \Lambda^\ast$ the coefficients
are the same as ones calculated
in Ref.~\cite{OctetEMFF} for
the elastic $\Lambda$ electromagnetic form factors.
As for the $\gamma^\ast \Sigma^0 \to \Lambda^\ast$
reaction, they are explicitly calculated, and
the coefficients are the same
as those for the reaction $\gamma^\ast \Lambda \to \Sigma^0$,
and reflect the isovector nature of the reaction~\cite{InPreparation}.
In both reactions, there is no interference between
the $\ket{M_A}_Y$ and $\ket{M_S}_Y$ states
which are in the initial and final states.

Note that, in the second term in Eq.~(\ref{eqSum})
there is a dependence on the diquark
polarization vectors $\varepsilon_{P_+}^\alpha (\lambda_D)$
and  $\varepsilon_{P_-}^{\beta \ast} (\lambda_D)$.
As already mentioned,
these states are defined according to
the fixed-axis representation~\cite{FixedAxis}
and depend also on the masses of the final
($M_{\Lambda^\ast}$) and initial ($M_Y$) states,
respectively.
Taking into account the sum in the diquark
polarization states we have~\cite{NDelta,FixedAxis},
\ba
\Delta^{\alpha \beta} & \equiv &\sum_{\lambda_D}
\varepsilon_{P_+}^\alpha (\lambda_D)
\varepsilon_{P_-}^{\beta\, \ast} (\lambda_D)  \nonumber \\
&=& -\left(g^{\alpha \beta} - \frac{P_-^\alpha P_-^\beta}{P_+ \cdot P_-}
\right) \nonumber \\
& &-
a \left( P_-- \frac{P_+ \cdot P_-}{M_{\Lambda^\ast}^2} P_+\right)^\alpha
\left( P_+- \frac{P_+ \cdot P_-}{M_Y^2} P_-\right)^\beta, \nonumber \\
& &
\label{eqDelta}
\ea
with
\be
a= \frac{M_{\Lambda^\ast} M_Y}{P_+ \cdot P_-
(M_{\Lambda^\ast} M_Y +P_+ \cdot P_-)}.
\ee

The calculation of the current~(\ref{eqJ1})
is carried out by the reduction from Eq.~(\ref{eqSum})
to the evaluation of a few matrix elements.
We present in Appendix~\ref{JY} the explicit expressions
for these matrix elements.

The final result is given by,
\ba
J_Y^\mu &=& + e \,\frac{1}{2}(3 j_1^A + j_1^S)
{\cal I}_Y
\, \hat \gamma^\mu  \gamma_5
\nonumber \\
& &- e\, \frac{1}{2}(3 j_2^A - j_2^S)
{\cal I}_Y \,
\frac{i \sigma^{\mu \nu} q_\nu}{2M}
  \gamma_5,
\label{eqJ2}
\ea
where
\ba
{\cal I}_Y (Q^2)= -
\eta_{\Lambda^\ast}
\int_k N (\varepsilon_0 \cdot \tilde k)
\psi_{\Lambda^\ast} (P_+,k)  \psi_Y (P_-,,k).
\label{eqInt0}
\ea
The integral ${\cal I}_Y$ is covariant and
includes the radial dependence of the wave functions.
We call ${\cal I}_Y$ the overlap integral.

\subsection{Form factors}

Combining Eqs.~(\ref{eqJS}) and  (\ref{eqJ2})
with the coefficients in Eq.~(\ref{eqJAS}), 
we obtain
the form factors
for the $\gamma^\ast \Lambda \to \Lambda^\ast$ reaction,
\ba
& &
F_1^\Lambda(Q^2)= \frac{1}{6}\left[f_{1+}(Q^2) -2 f_{10}(Q^2)\right]
{\cal I}_\Lambda,
\label{eqF1}\\
& &
F_2^\Lambda(Q^2)= \frac{1}{3} f_{20}(Q^2)
\frac{M_{\Lambda^\ast} +M_\Lambda}{2M}
{\cal I}_\Lambda.
\label{eqF2}
\ea
As for the $\gamma^\ast \Sigma^0 \to \Lambda^\ast$ reaction,
using Eq.~(\ref{eqJAS2}) and we obtain:
\ba
& &
F_1^\Sigma(Q^2)= -\frac{1}{\sqrt{12}} f_{1-}(Q^2)
{\cal I}_\Sigma,
\label{eqF1s}\\
& &
F_2^\Sigma(Q^2)= + \frac{2}{\sqrt{12}} f_{2-}(Q^2)
\frac{M_{\Lambda^\ast} +M_\Sigma}{2M}
{\cal I}_\Sigma.
\label{eqF2s}
\ea
Note that the presence of the factor $2M$
in the form factor expressions, which is a consequence
of the quark Pauli current expressed in terms of the nucleon
mass $M$~\cite{Nucleon,Omega} in Eq.~(\ref{eqJq}).

The overlap integral ${\cal I}_Y$ can be evaluated
in the $\Lambda^\ast$ rest frame to give
a simple expression~\cite{S11},
\ba
{\cal I}_Y (Q^2) =
 \eta_{\Lambda^\ast}
\int_k \frac{k_z}{|{\bf k}|}
\psi_{\Lambda^\ast} (P_+,k)
\psi_Y(P_-,k),
\label{eqInt2}
\ea
where
\ba
P_-=(E_Y,0,0,-|{\bf q}|_Y), \hspace{.5cm}
P_+=(M_{\Lambda^\ast},0,0,0),
\label{eqLRF}
\ea
with $E_Y=\sqrt{M_Y^2 + |{\bf q}|_Y^2}=
\sfrac{M_{\Lambda^\ast}^2 + M_Y^2 + Q^2}{2 M_{\Lambda^\ast}}$.

From Eq.~(\ref{eqInt2}) we may conclude
that the signs of the overlap integrals for $Y=\Lambda$
and $Y=\Sigma$ depend on the relative sign
of the $\Lambda$ and $\Sigma$ scalar wave functions.
Defining the factor,
\ba
\eta_{\Lambda \Sigma}= \frac{{\cal N}_\Lambda {\cal N}_\Sigma}
{|{\cal N}_\Lambda {\cal N}_\Sigma|},
\ea
which gives the relative sign between
the $\Lambda$ and $\Sigma$ radial wave functions,
we can write in the limit $M_\Lambda = M_{\Sigma^0}$ as,
\ba
\mbox{sgn} ({\cal I}_{\Sigma})=
\eta_{\Lambda \Sigma^0}  \times
\mbox{sgn} ({\cal I}_{\Lambda}).
\label{eqSign}
\ea
This result is equivalent to state
that the relative sign of the integrals
${\cal I}_\Lambda$ and ${\cal I}_{\Sigma}$
is given by the relative
sign of ${\cal N}_\Lambda$ and ${\cal N}_\Sigma$
(or $\eta_{\Lambda \Sigma^0}$).
Since $M_\Lambda$ and $M_{\Sigma^0}$ values are close,
it is expected that the relation~(\ref{eqSign})
holds also for a certain region of $Q^2$.
The phase $\eta_{\Lambda \Sigma^0}$ is unknown at present,
as the same reason for the sign of the
$\gamma^\ast \Lambda \to \Sigma^0$
transition magnetic moment $\mu_{\Lambda \Sigma^0}$
is unknown~\cite{PDG}.
If the sign of $\mu_{\Lambda \Sigma^0}$ is determined,
we may be able to fix the sign for the
$\gamma^\ast \Sigma^0 \to \Lambda^\ast$
transition form factors within the present approach.
Therefore, although we will assume $\eta_{\Lambda \Sigma^0}=1$
in the presentation of our results later,
we will also discuss the alternative sign possibility.

For later discussions, it is also important to mention
that the integral~(\ref{eqInt0}) has a behavior,
\ba
{\cal I}_Y(Q^2) \propto |{\bf q}|_Y,
\label{eqIntProp}
\ea
for small $|{\bf q}|_Y$.
Recall that $|{\bf q}|_Y$, given by Eq.~(\ref{eqqY}),
is the photon three-momentum in the $\gamma^\ast Y \to \Lambda^\ast$ reaction
in the final $\Lambda^\ast$ rest frame.
See Appendix C of Ref.~\cite{S11} for
the derivation of the relation~(\ref{eqIntProp}).

We can now discuss the $Q^2$ range applicable
for the present model.
From the definition of the transition form factors~(\ref{eqJS}),
we can conclude that the Dirac-type form factor $F_1^Y$
should be zero or vanish when $Q^2 \to 0$.
However, in the present case if $M_{\Lambda^\ast} \neq M_Y$, 
it results to $F_1^\ast(0) \ne 0$.
That is a simple consequence  of the relation~(\ref{eqIntProp}),
from what we can conclude that ${\cal I}_Y(0) \ne 0$
in the case $Q^2=0$,
when $|{\bf q}|_Y=
|{\bf q}|_{0Y}=\sfrac{M_{\Lambda^\ast}^2 -M_Y^2}{2 M_{\Lambda^\ast}}$.
This result is equivalent with that the
$Y$ and $\Lambda^\ast$ states are not orthogonal
in the spectator quark model\footnote{
This is a consequence of the fact
that we cannot have simultaneously the $Y$ and $\Lambda^\ast$
at rest when $Q^2=0$, unless the particles have the same masses.
Considering for instance $\Lambda^\ast$ at rest in the following.
According to Eq.~(\ref{eqLRF}), one gets
$P_+=(M_{\Lambda^\ast},0,0,0)$, but
\mbox{$P_-= \left( \sfrac{M_{\Lambda^\ast}^2+ M_Y^2}{2 M_{\Lambda^\ast}},0,0,
-|{\bf q}|_{0Y}\right)$.}
Therefore $Y$ is not at rest.}.
The two states would be orthogonal
only in the case $M_{\Lambda^\ast}=M_Y$, when $|{\bf q}|_{0Y}=0$.
We can regard the states
are approximately orthogonal
when $|{\bf  q}|_{0Y}$ is very small, which
leads to ${\cal I}_Y(0) \approx 0$.
Then, we can assume that the condition ${\cal I}_Y(0) \simeq 0$
is satisfied when $Q^2 \gg |{\bf q}|_{0Y}^2$.

Interpreting $\Lambda^\ast$ as $\Lambda(1670)$
($M_{\Lambda^\ast} \simeq 1.670$ GeV)
the model is then applicable when $Q^2 \gg |{\bf q}|_0^2 = 0.21$ GeV$^2$
($M_\Lambda=1.116$ GeV) for the reaction involving the $\Lambda$,
and when $Q^2 \gg |{\bf q}|_0^2 = 0.17$ GeV$^2$
($M_{\Sigma^0}=1.193$ GeV) for the reaction involving the $\Sigma^0$.

\section{Chiral unitary model}
\label{secCUM}

In this section, we briefly explain the description
of the $\Lambda(1670)$ resonance
and the calculation of the corresponding form factors in the chiral unitary approach.
Here we consider the model
presented in Refs.~\cite{Oset:2001cn,Doring10}.

\subsection{Description of $\Lambda(1670)$}

In the chiral unitary model, the $\Lambda(1670)$ is dynamically generated in $s$-wave meson-baryon scattering in the coupled channels of $\bar KN$, $\pi\Sigma$, $\eta\Lambda$, $K\Xi$, $\pi\Lambda$ and $\eta\Sigma$ with
zero total charge.
Here we take small isospin breaking into account in the masses of the mesons and baryons.
The $s$-wave scattering amplitude in these channels is calculated with the scattering equation given by
\begin{equation}
   T(W) = V(W) + V(W) G(W) T(W),
\end{equation}
where $W$ is the center of mass energy of the two-body system.
Based on the $N/D$ method with neglecting the left-hand cut, a solution of the scattering equation can be obtained by a simple algebraic equation~\cite{Oller:2000fj}
\begin{equation}
   T = (1-VG)^{-1} V . \label{eq:Tmat}
\end{equation}

For the interaction kernel $V$ in Eq.~\eqref{eq:Tmat} we take the lowest order of the chiral perturbation theory, which is the Weinberg-Tomozawa term, as
\begin{equation}
   V_{ij} = - C_{ij} \frac{1}{4f^{2}} (2W - M_{i}- M_{j}) N_{i} N_{j}, \label{eq:WTterm}
\end{equation}
with the coupling strength $C_{ij}$, the meson decay constant $f$ being fixed as $f=1.123 f_{\pi}$ with $f_{\pi}=93$ MeV, the baryon mass
$M_{i}$ and the normalization of baryon state $N_{i} \equiv \sqrt{(M_{i}+E_{i})/(2M_{i})}$
where $E_{i}$ is the baryon energy in the c.m.~frame. It is important to note that the coupling strength $C_{ij}$ is fixed solely by the flavor $SU(3)$ group structure of the
channel, and thus once we fix the meson decay constant, there are no free parameters
in the interaction \eqref{eq:WTterm}. We do not include an explicit pole term in the
interaction. This is the reason that the obtained resonance in the scattering
amplitude is called a dynamically generated resonance.

The diagonal matrix $G$ in Eq.~\eqref{eq:Tmat} is the meson-baryon loop function given by
\begin{equation}
   G_{i}(W) = i \int \frac{d^{4}p}{(2\pi)^{4}} \frac{2M_i}{p^{2} - M_{i}^{2} + i\epsilon } \frac{1}{(P - p)^{2} - m^{2}_{i} + i \epsilon},
\end{equation}
with the c.m.\ energy $P=(W,0,0,0)$ and the meson mass $m_{i}$.
The divergent loop function can be calculated in an analytic form using
dimensional regularization, which isolates the divergent part from the integral.
The remaining finite constant being called $a_{i}$ is determined phenomenologically
by experiments. Here we use the threshold branching ratios of $K^{-}p$ to $\pi\Lambda$
and $\pi\Sigma$ observed by stopped $K^{-}$ mesons in
hydrogen~\cite{Tovee:1971ga,Nowak:1978au}. In this study use the following $a_{i}$
constants determined in Ref.~\cite{Oset:2001cn}:
\begin{equation}
  \begin{array}{ccc}
   a_{\bar K N} = -1.84, \hspace{0.3cm}& a_{\pi \Sigma} = - 2.00,
   \hspace{0.3cm}& a_{\pi \Lambda} = -1.83, \\
   a_{\eta \Lambda} = -2.25, \hspace{0.3cm} & a_{\eta \Sigma} = -2.38,
   \hspace{0.3cm} & a_{K \Xi} = -2.67,
   \end{array}
   \label{subtsorig}
\end{equation}
with the scale of the dimensional regularization $\mu =630$ MeV.

Since the obtained amplitude is written in an analytic form,
we can perform analytic continuation to the complex energy plain
to look for resonances poles in the
second Riemann sheet. The pole position for the $\Lambda(1670)$ resonance
in this model can be found at
\begin{equation}
   z = 1680 - 20 i \ {\rm [MeV]}. \label{eq:pole}
\end{equation}
We also obtain the coupling strength $g_{\Lambda^{*}}^{i}$
of the $\Lambda(1670)$ to the channel $i$ as a residue of the scattering amplitude
at the resonance pole. The values of the couplings are given in
Ref.~\cite{Doring10}. The couplings characterize the structure of the $\Lambda(1670)$.
The $\Lambda(1670)$ has large couplings to the $\eta \Lambda$ and $K\Xi$ channels.
As discussed in Ref.~\cite{Hyodo:2008xr}
the values of the constants $a_{i}$ are very important for the nature of the dynamically
generated resonance.
If we take the constants $a_{i}$ determined in the natural renormalization
scheme which excludes the CDD pole contributions~\cite{Hyodo:2008xr},
we obtain a resonance pole at $1700 - 21i$ MeV~\cite{Doring10}. This is not so different
from the pole position \eqref{eq:pole} determined phenomenologically by the $K^{-}p$
threshold branching ratios.
This means that the resonance obtained in this parameter
set is composed mostly by meson-baryon components.

\subsection{Transition amplitude}

We calculate the transition amplitude of the
$\Lambda(1670)$ resonance
using the method developed in Ref.~\cite{Jido08}.
In the following we adopt an alternative
parametrization for the transition current to Eq.~(\ref{eqJS})
as given in Ref.~\cite{Doring10}:
\begin{equation}
  J_{Y,\rm NR}^{\mu}  = {\cal M}_{1}^{\rm NR} \sigma^{\mu}
+ {\cal M}^{\rm NR}_{2} P^{\mu}_{+} \sigma\cdot
  q + {\cal M}^{\rm NR}_{3} q^{\mu} \sigma \cdot q  , \label{eq:nonreladecomp}
\end{equation}
where $\sigma^{\mu}=(0,{\boldsymbol \sigma})$ with the Pauli matrix $\sigma^{i}$ for the
hyperon spin space and $P^{\mu}_{+} = (M_{\Lambda^\ast},0,0,0)$ and $q^{\mu}$ are the $\Lambda^{*}$
and photon momenta, respectively.
The current $J_{Y,\rm NR}^{\mu}$ is projected
on the $Y$ and $\Sigma^0$ Pauli spinors.
This representation is equivalent to the transition current
$J_Y^\mu$ of Eq.~(\ref{eqJS})
once one understands that the spin projection on
the asymptotic state Dirac spinors $u_Y(P_-,S_z)$
and $\bar u_{\Lambda^\ast}(P_+, S_z^\prime)$
is already performed in the $\Lambda^\ast$ rest frame~(\ref{eqLRF}).
The index ${\rm NR}$ is intend to indicate that
we will make a non-relativistic reduction of the operators and
take the leading order contributions,
but still the current itself is covariant.
The parametrization (\ref{eq:nonreladecomp}) together
with the gauge invariance condition,
\begin{equation}
  {\cal M}_{1}^{\rm NR}  + {\cal M}^{\rm NR}_{2} q\cdot P_{+} + {\cal M}^{\rm NR}_{3} q^{2}  = 0, \label{eq:gauseinvNR}
\end{equation}
is equivalent to the representation of Eq.~(\ref{eqJS}).

With these amplitudes the transition form factors are written as
\begin{eqnarray}
   F_{1}^{Y}(Q^{2})
   &=&
    Q^{2} \frac{1}{e}\sqrt{\frac{1}{1+\tau}} \sqrt{\frac{M_{Y}}{M_{\Lambda^{*}}}}
    \nonumber \\ && \times
    \left( \frac{M_{\Lambda^{*}}}{M_{Y}+M_{\Lambda^{*}}} {\cal M}_{2}^{\rm NR} + {\cal M}_{3}^{\rm NR} \right),
 \label{eq:F1ChUM}
  \\
  F_{2}^{Y}(Q^{2})
  &=&
  (M_{Y}+M_{\Lambda^{*}})^{2}\frac{1}{e} \sqrt{\frac{1}{1+\tau}}  \sqrt{\frac{M_{Y}}{M_{\Lambda^{*}}}}
 \nonumber \\ && \times
 \left( - \frac{M_{\Lambda^{*}}}{M_{Y}+M_{\Lambda^{*}}} {\cal M}_{2}^{\rm NR} + \tau {\cal M}_{3}^{\rm NR} \right),
\label{eq:F2ChUM}
\end{eqnarray}
where
$\tau$ is given by Eq.~\eqref{eqtau},
$M_{Y}$ and $M_{\Lambda^{*}}$ are the masses of the hyperon $Y$
and $\Lambda^{*}$, respectively,
and we set\footnote{In the previous work~\cite{Doring10}, $M_{\Lambda^{*}}=1680$ MeV
was used. This value corresponds to the real part of the pole position for
the $\Lambda(1670)$ in the present model.} $M_{\Lambda^{*}}=1670$ MeV.
In the above equations,
the factor $\sfrac{1}{e}$ must be included
since the form factors defined by (\ref{eqJS})
are defined without $e$
and the transition amplitudes ${\cal M}_i^{\rm NR}$
include the factor $e$ as shown next.
The absolute phases of $F_{1}^{Y}$ and $F_{2}^{Y}$ are arbitrary in the present model.
Here we define the phases of the transition form factors obtained in the chiral unitary
model so that the value of $A_{1/2}^{Y}(Q^{2})$ at $Q^{2}=0$ for each hyperon
$Y$ should be real and positive. This is equivalent to set the value of $F_{2}^{Y}(0)$
real and negative from Eq.~\eqref{eqA12X} with $F_{1}^{Y}(0)=0$ thanks to
gauge invariance.

\begin{figure}
\includegraphics[width=0.47\textwidth]{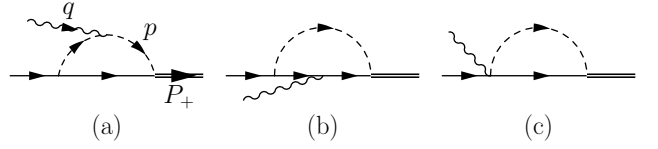}
\caption{Feynman diagrams for the phototransition to the $\Lambda^{*}$. The solid, dashed, wavy and double lines denote octet baryons, mesons, photon and $\Lambda^{*}$, respectively. The diagram~(b) gives sub-leading contribution in the nonrelativistic limit. \label{fig:FeynmanDiagram}}
  \label{fig:loops}
\end{figure}

The transition amplitudes ${\cal M}_{i}^{\rm NR}$ are calculated
based on the Feynman diagrams shown in Fig.~\ref{fig:FeynmanDiagram}
in a nonrelativistic formulation in which the operators are expanded in terms of
$1/M_{i}$ and only the leading contributions are taken.
The amplitudes are decomposed in terms of the Lorentz structures given
by Eq.~(\ref{eq:nonreladecomp}). As shown in  Eqs.~(\ref{eq:F1ChUM})
and~(\ref{eq:F2ChUM}),
the transition form factors can be expressed by ${\cal M}_{2}^{\rm NR}$ and ${\cal M}_{3}^{\rm NR}$.
Since it was found in Ref.~\cite{Jido08} that the diagram (c) has only ${\cal M}_{1}^{\rm NR}$ term,
which is irrelevant for the form factors, we can actually omit the diagram (c) for the present purpose.
It should be noted that the amplitudes ${\cal M}_{2}^{\rm NR}$ and ${\cal M}_{3}^{\rm NR}$
remain finite although each process contains one-loop integral.
Since the diagram (b) has the $\gamma BB$ vertex having the $1/M_{i}$ factor,
the diagram (b) gives only sub-leading contribution
in the nonrelativistic limit and we neglect the diagram (b).

Each vertex in the
diagrams is given by the chiral effective theory.
The basic interactions of the mesons and baryons are given by the chiral Lagrangian:
\begin{eqnarray}
{\cal L}_{MBB} &=&
 - \frac{D}{\sqrt 2 f} \, {\rm Tr} \left[ \bar B \gamma_{\mu} \gamma_{5} \{\partial^{\mu}\Phi,B\} \right]
 \nonumber \\ &&
 - \frac{F}{\sqrt 2 f}\, {\rm Tr} \left[ \bar B \gamma_{\mu} \gamma_{5} [\partial^{\mu}\Phi ,B] \right] \label{eq:MBcoup},
\end{eqnarray}
with the meson and baryon fields, $\Phi$ and $B$, defined by
\begin{eqnarray}
   \Phi &=&
    \left(
   \begin{array}{ccc}
       \frac{1}{\sqrt{2}} \pi^0 + \frac{1}{\sqrt{6}} \eta & \pi^+ & K^+ \\
       \pi^- & -\frac{1}{\sqrt{2}} \pi^0 + \frac{1}{\sqrt{6}} \eta  & K^0\\
       K^- & \bar K^0 & - \frac{2}{\sqrt{6}} \eta
   \end{array}
   \right),
\\
   B &=&
   \left(
   \begin{array}{ccc}
       \frac{1}{\sqrt{2}} \Sigma^0 + \frac{1}{\sqrt{6}} \Lambda &
       \Sigma^+ & p \\
       \Sigma^- & -\frac{1}{\sqrt{2}} \Sigma^0 +
       \frac{1}{\sqrt{6}} \Lambda  & n\\
       \Xi^- & \Xi^0 &- \frac{2}{\sqrt{6}} \Lambda
   \end{array}
   \right). \ .
\end{eqnarray}
The meson-baryon coupling constants are obtained from the Lagrangian as
$g_{A}^{i}/(2f)$ with the axial coupling constant $g_{A}^{i}$ given by $D$ and $F$
together with the Clebsch-Gordan coefficients.
The parameters are fixed as
\begin{equation}
D=0.85 \pm 0.06 \ ,\ \ \ \ \ F=0.52 \pm 0.04 \ ,  \label{eq:DFvalues}
\end{equation}
so as to reproduce the observed axial vector coupling for the octet baryons.
The photon couplings to mesons and baryons are given by the gauge coupling:
\begin{eqnarray}
    {\cal L}_{\gamma B} &=& -e {\rm Tr}\left[ \bar B \gamma_{\mu} [Q_{\rm ch},B]\right] A^{\mu}, \label{eq:LaggBB} \\
    {\cal L}_{\gamma M} &=& ie {\rm Tr}\left[ \partial_{\mu} \Phi [Q_{\rm ch},\Phi]\right] A^{\mu},
\end{eqnarray}
with the charge matrix $Q_{\rm ch} = {\rm diag}(\frac{2}{3},-\frac{1}{3},-\frac{1}{3})$
and $e>0$. The Kroll-Ruderman terms of the $\gamma MBB$ couplings are
obtained by replacing the derivative acting on the meson fields $\partial_{\mu} \Phi$
with the covariant derivative $D_{\mu} \Phi = \partial_{\mu} \Phi + i e A_{\mu}[ Q_{\rm ch}, \Phi]$ in the Lagrangian~\eqref{eq:MBcoup}. The $\Lambda^{*}$ coupling to the meson and baryon has an $s$-wave form
\begin{equation}
   {\cal L}_{\Lambda^{*}M_{i}B_{i}} = g_{\Lambda^{*}}^{i} \bar \Lambda^{*} \Phi_{i} B_{i} \ ,
   \label{effective}
\end{equation}
with the coupling constant $g_{\Lambda^{*}}^i$ determined
by the chiral unitary model. The explicit values are given in Ref.~\cite{Doring10}.

The amplitude given by $-it=J\cdot \epsilon$ for the diagram (a) with channel $i$ is calculated as
\begin{eqnarray}
   -it_{a}^{i}  & = &
   iQ_{M} A_{i}   \int \frac{d^{4} p}{(2\pi)^{4}}
   \frac{  ( p - q) \cdot \sigma  \,   (2p-q)\cdot \epsilon}
   {(P_{+}-p)^{2} - M^{2}_{i}+ i\epsilon}\nonumber \\
   &&\times\frac{1}{(p^{2} - m^{2}_{i} + i\epsilon)((p-q)^{2} - m^{2}_{i}+ i\epsilon )},
\end{eqnarray}
with $Q_M$ the meson ($M$) charge and $A_{i}$ is given by
$A_{i}= g_{A}^{i} g_{\Lambda^{*}}^{i} M_{i} / f $.
After some algebra shown in Ref.~\cite{Jido08,Doring10}, we obtain
\begin{eqnarray}
   -it_{a}^{i}   & = &
 iQ_{M} A_{i} 2 \int^{1}_{0} dx \int^{x}_{0} dy
   \int \frac{d^{4} p}{(2\pi)^{4}}
   \frac{  ( p +(y-1) q) \cdot \sigma }
   { \left(p^{2} - S_{a}^{i} + i\epsilon \right)^{3}} \nonumber \\
&&\times   \left(2p+(2y-1)q +2(1-x) P_{+}\right)\cdot \epsilon
   \ , \label{eq:diaaNR}
\end{eqnarray}
where $S_{a}^{i}$ is defined by
\begin{eqnarray}
   S_{a}^{i} &=&  2 P_{+}\cdot q (1-x)y -
M_{\Lambda^{*}}^2 x(1-x) -  q^{2} y (1-y)   \nonumber  \\ &&
   + M^{2}_{i} (1-x) + m_{i}^{2}x \ . \label{eq:defSa}
\end{eqnarray}
In Eq.~(\ref{eq:diaaNR}), only even powers of $p$ give contribution after
performing  the integration.  The ${\cal M}^{\rm NR}_{2}$ and
${\cal M}^{\rm NR}_{3}$ amplitudes can be calculated as finite numbers.
After performing the integration,  we get the ${\cal M}^{\rm NR}_{2}$ and ${\cal M}^{\rm NR}_{3}$ components  for the channel $i$ as
\begin{eqnarray}
    {\cal M}^{i \rm (NR)}_{2a} &=&    \frac{ Q_{M} A_{i}}{(4\pi)^{2}}
    \int^{1}_{0} dx \int^{x}_{0} dy  \frac{2(y-1)(1-x)}{S_{a}^{i} - i\epsilon }, \label{eq:amp2NR}
    \\
    {\cal M}^{i \rm (NR)}_{3a} &=&   \frac{ Q_{M} A_{i}}{(4\pi)^{2}}
    \int^{1}_{0} dx \int^{x}_{0} dy  \frac{(y-1)(2y-1)}{S_{a}^{i} - i\epsilon }. \label{eq:amp3NR}
\end{eqnarray}

In order to take into account the charge distribution of the constituent mesons
and baryons, we multiply the transition amplitudes obtained above by the electromagnetic form factors of the mesons or baryons to which the photon couples.
The $Q^2$ dependence of the helicity amplitude of  the
$\Lambda^{*}$ resonance, thus, stems from the form factors  of the meson and
baryons components and the intrinsic  $Q^2$ structure of the loops. For the
mesons and baryons form factors, we  take monopole form factors:
\begin{equation}
F(Q^2) = \frac{\Lambda^2}{\Lambda^2+ Q^2}, \label{eq:FormFactor}
\end{equation}
with
\begin{eqnarray}
\Lambda_\pi &=& 0.727\ {\rm [GeV]}, \\
\Lambda_K &=& 0.828\ {\rm [GeV]},
\end{eqnarray}
which are determined by the radii of the mesons.
These values correspond to $\langle r^2 \rangle=0.44$ fm$^2$ and
$\langle r^2 \rangle=0.34$ fm$^2$ for the pion and the kaon,  respectively.
For the baryon, we take the same form factor as for
the corresponding meson to keep  gauge invariance.
Thanks to the practically negligible
effect of the baryon terms,
the approximation made there has no practical consequences.

\section{Results}
\label{secResults}

We first present the results of the
valence quarks (spectator quark model)
and meson cloud (chiral unitary model)
for the $\gamma^\ast N \to N^\ast(1535)$
transition form factors in the charge $+1$ channel (namely for the proton target case).
Although some of the results were already reported
in the previous works~\cite{Jido08,S11},
we present again some results of
the form factors, since they are important
and can make the later discussions clearer.

After analyzing the results for the $\gamma^\ast N \to N^\ast(1535)$
reaction, we will discuss the reaction $\gamma^\ast Y \to \Lambda^\ast$
for $Y=\Lambda$ and $\Sigma^0$.
The results of the $\gamma^\ast Y \to \Lambda^\ast$ reactions
will be compared with those of the $\gamma^\ast N \to N^\ast(1535)$.

We recall that the applicable region of the
present valence quark model is $Q^2 \gtrsim 1$ GeV$^2$.
As for the chiral unitary model,
we cannot extend the results for an arbitrary large $Q^2$,
because the amplitudes are calculated
using the vertex given by the chiral perturbation theory.
Therefore, we expect the results
of both formalisms can be compared
in the region $Q^2=1-2$ GeV$^2$,
where the correlation between the two effects
can possibly determine the final result for
the transition form factors.

About the chiral unitary model
we recall that the contributions
from the valence quarks (or baryon core)
for the form factors are real numbers.
The results from the chiral unitary model
are based on a meson-baryon coupled-channels formalism~\cite{Jido08,Doring10}
and the states are constructed as a consequence of the
meson cloud dressing of the bare octet baryons,
and the transition amplitudes are calculated by photon couplings to the hadron constituents.
In the diagrams with the baryon dressing
(see Fig.~\ref{fig:FeynmanDiagram})
one can have on-mass-shell states
for the mesons or baryons,
therefore the amplitudes and the
form factors become complex number functions.
As we have already mentioned, the absolute phase is fixed so as to make
$A_{1/2}$ to be real and positive, or equivalently $F_{2}$ real and negative,
at $Q^{2}=0$.

\subsection{$\gamma^\ast N \to N^\ast(1535)$ form factors}

\begin{figure}[t]
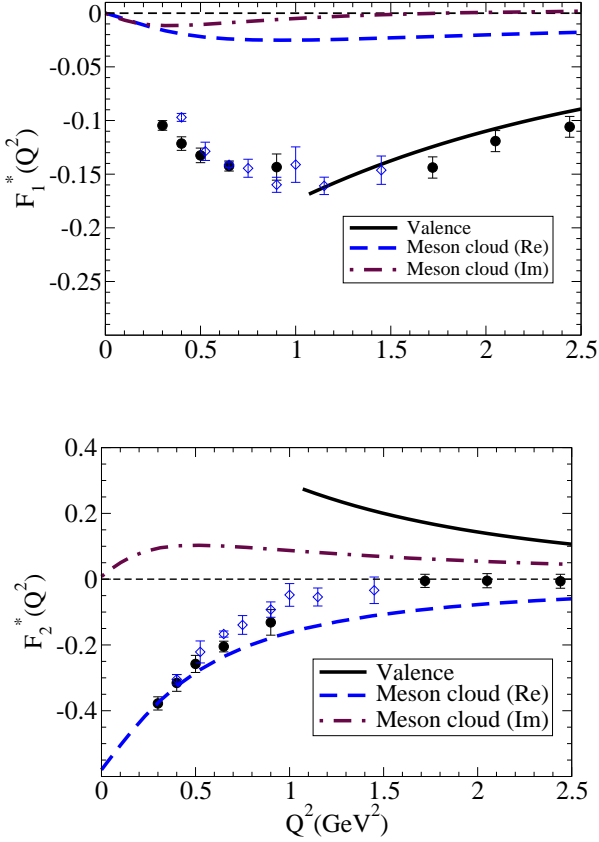

\vspace{.3cm}
\centerline{
\includegraphics[width=3.1in]{F1_S11x.eps} }
\vspace{1.cm}
\centerline{
\includegraphics[width=3.0in]{F2_S11.eps} }
\caption{\footnotesize{ (Color online)
Valence quark and meson cloud contributions
for the $\gamma^\ast N \to N^\ast(1535)$ transition form factors,
$F_1^\ast(Q^2)$ and $F_2^\ast(Q^2)$.
While the valence quark contributions are obtained by the
covariant spectator quark model~\cite{S11},
those of the meson cloud contributions are obtained by
the chiral unitary model~\cite{Jido08}.
Data are from CLAS~\cite{CLAS} and MAID~\cite{MAID}.
}}
\label{figS11}
\end{figure}

The results of the covariant spectator quark model
and the chiral unitary model for the
$\gamma^\ast N \to N^\ast(1535)$ transition
form factors are presented in Fig.~\ref{figS11}.
The individual results for the helicity amplitudes
were presented in Ref.~\cite{S11} (for the valence quarks)
and in Ref.~\cite{Jido08} (for the meson cloud).
The results from the covariant spectator quark model are restricted
to the region $Q^2> 1$ GeV$^2$, since the applicability of the model
requires $Q^2 \gg \left( \sfrac{M_S^2- M^2}{2M_S} \right)^2
\simeq 0.21$ GeV$^2$, where $M_S$ corresponds to this case
to the $N^\ast(1535)$ mass~\cite{S11}.

For the $\gamma^\ast N \to N^\ast(1535)$ reaction
we can observe different roles of the
valence quark and meson cloud degrees of freedom.
Since in both cases for $F_1^\ast$ and $F_2^\ast$
the imaginary part is small,
we will focus only on the real part.
In Fig.~\ref{figS11} one can notice the dominance
of the valence quark effect for the Dirac-type form factor $F_1^\ast$,
with the prediction very close to the data~\cite{CLAS,MAID}
for $Q^2 > 1$ GeV$^2$.
In this case the meson cloud contributions
are about an order of magnitude smaller than
those of the valence quarks.
As for the Pauli-type form factor $F_2^\ast$,
one can see, on the other hand, that
the meson cloud contributions
are sufficient to explain the data for $Q^2 < 1$ GeV$^2$.
Furthermore, the valence and meson cloud contributions
have opposite signs with similar magnitude
for $Q^2 > 1$ GeV$^2$.
The cancellation between the
two contributions may be the main reason
of the experimental result, $F_2^\ast \simeq 0$
for $Q^2 > 1$ GeV$^2$ (see $F_2^\ast$ in Fig.~\ref{figS11}).

For later convenience, we study also
the falloff of the form factors for the valence quark
contributions in the large $Q^2$ region.
Apart from logarithm corrections
(very smooth variation with $Q^2$)~\cite{S11},
the falloff behavior of the form factors
can be expressed
by $F_1^\ast(Q^2) \approx \left(\sfrac{\Lambda_1^2}{\Lambda_1^2+Q^2}
\right)^{2}$
and
\mbox{$F_2^\ast(Q^2) \approx \left(\sfrac{\Lambda_2^2}{\Lambda_2^2+ Q^2}
\right)^{3}$},
where $\Lambda_1^2 \simeq 2.6$ GeV$^2$ and
$\Lambda_2^2 \simeq 2.7$ GeV$^2$ for $Q^2 \simeq 2$ GeV$^2$.
Therefore, the $\gamma^\ast N \to N^\ast(1535)$
transition form factors have much slower falloff
than that for the nucleon elastic form factors,
where the corresponding cutoff is $\Lambda^2 = 0.71$ GeV$^2$.

We recall that the individual contributions,
those from the valence quarks and meson cloud,
are based on the different frameworks.
The valence quark contributions are estimated
by a constituent quark model that takes into account
the quark internal electromagnetic structure
(including possible quark-antiquark internal excitations),
but it does not include the processes where
a meson is created by the overall baryon.
On the other hand, the meson cloud contributions
are estimated in the meson-baryon interactions
where both states are considered as
structureless particles but modified by
monopole meson form factors (see Sec.~\ref{secCUM}).
Due to the differences in
the degrees of freedom used in the
two approaches described above, we cannot
simply combine the individual contributions
to get total results for the form factors.
However, the opposite signs of the individual
contributions for $F_2^\ast$ are very suggestive,
that a strong cancellation between
the valence quark and meson cloud effects
may take place in a unified approach.

The results shown for the $\gamma^\ast N \to N(1535)$
reaction suggest that
the form factor representation may be very
convenient to analyze the
transition between the nucleon and
the first excited state of the nucleon with a negative parity.
Using the form factor representation,
it is clear that while $F_1^\ast$ is dominated by the valence
quark contributions, $F_2^\ast$ may be a result of the
competition between the valence quark and meson cloud effects.
This simple separation is not obvious
in the helicity amplitude representation.

The results obtained for the
$\gamma^\ast N \to N^\ast(1535)$ reaction,
and the simplified interpretation
in terms of the individual (valence quarks
and meson cloud) contributions,
raise a question, namely, whether or not such a trend
can be observed for 
similar reactions.
Therefore, we next study the
$\gamma^\ast \Lambda \to \Lambda^\ast$ ($Y=\Lambda, \Sigma^0$)
reactions with $\Lambda^\ast = \Lambda(1670)$.

\subsection{$\gamma^\ast \Lambda \to \Lambda^\ast$
and $\gamma^\ast \Sigma^0 \to \Lambda^\ast$
transition form factors}

\begin{figure*}[t]
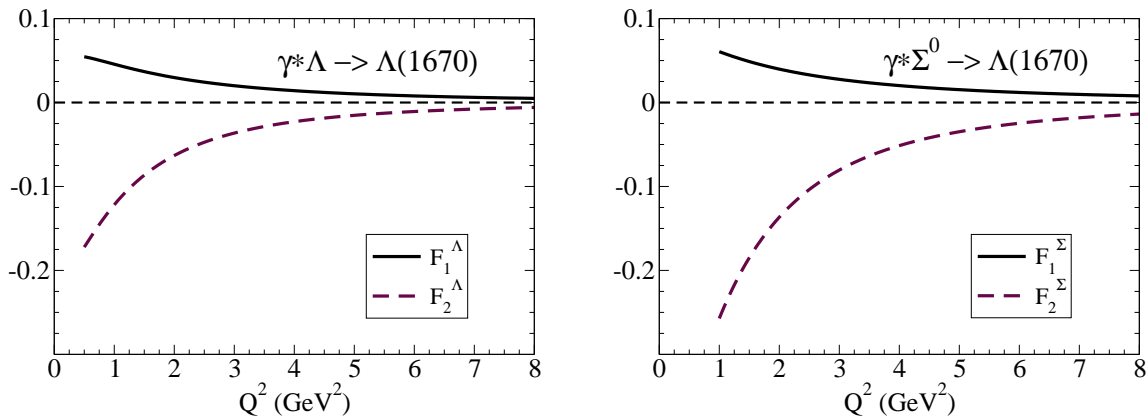

\centerline{
\mbox{
\includegraphics[width=2.8in]{F1F2total.eps} }
\hspace{.5cm}
\mbox{
\includegraphics[width=2.8in]{F1F2SigmaT.eps} }}
\vspace{.1cm}
\caption{\footnotesize{ (Color online)
Valence quark contributions for the
$\gamma^\ast Y \to \Lambda(1670)$
form factors, for $Y=\Lambda$ (left panel), and
$Y=\Sigma^0$ (right panel).}}
\label{figF1F2}
\end{figure*}

We now discuss the
$\gamma^\ast Y \to \Lambda^\ast$ reactions,
for $Y=\Lambda$ and $\Sigma^0$.
As in the previous section we will
compare the contributions from the valence quarks
and those from the meson cloud dressing
for the corresponding form factors.
The results of the valence quark contributions
derived from the covariant spectator quark model
are given in Sec.~\ref{secSQM}
[Eqs.~(\ref{eqF1})-(\ref{eqF2s})].
The meson cloud contributions
calculated in Ref.~\cite{Doring10}
using the chiral unitary model,
are reviewed in Sec.~\ref{secCUM}
[Eqs.~(\ref{eq:F1ChUM}) and~(\ref{eq:F2ChUM})].

\subsubsection{Results of spectator quark model}

First, we discuss the valence quark contributions,
which are presented here
for the first time, using the
covariant spectator quark model.
As mentioned already, the valence quark
contributions depend on the two different phases (signs),
$\eta_{\Lambda^\ast}$, the relative sign between the $\Lambda$
and $\Lambda^\ast$ states,
and $\eta_{\Lambda \Sigma^0}$ given
by the relative sign between the
$\Lambda$ and $\Sigma^0$ radial wave function
normalization constants.
We first consider $\eta_{\Lambda^\ast}=1$ case,
since it is equivalent to the
phase for the $N^\ast(1535)$-nucleon case
as already discussed.
The sign was determined by the experimental form factor
data\footnote{The sign
of the $N^\ast(1535)$ wave function
was adjusted to generate $F_1^\ast(Q^2) < 0$,
in agreement with the data
from Refs.~\cite{CLAS,MAID}.
In fact, for $F_1^\ast$ the
valence quark contributions give
a very good approximation to describe
the data~\cite{S11,S11scaling}.
}~\cite{S11} (see Fig.~\ref{figS11}).
As for the reaction involving the $\Sigma^0$,
we take $\eta_{\Lambda \Sigma^0} =1$,
which is equivalent to state that the $\Lambda$ and $\Sigma$
radial wave function normalization constants are both positive.

The results are presented in Fig.~\ref{figF1F2}.
The $Q^2$ region shown is extended up to $Q^2=8$ GeV$^2$,
in order to observe better the falloff behavior of the
$F_1^Y$ and $F_2^Y$ taking advantage of the covariant
nature of the model.
We focus on the region $Q^2 > 1$ GeV$^2$
that satisfies the model applicable condition,
$Q^2 \gg 0.2$ GeV$^2$ for both the reactions,
since $|{\bf q}|_{0Y}^2 \approx 0.2$ GeV$^2$.
We can observe in Fig.~\ref{figF1F2} the slow falloff
for the both form factors in both reactions, particularly
for $F_1^Y$.
We will come back later to the falloff of the form factors.

Another interesting point in Fig.~\ref{figF1F2}
is the magnitude of the form factors
$F_1^Y$ and $F_2^Y$ which is very similar
for the both cases $\Lambda$ and $\Sigma^0$.
However, the form factors for
the reaction with $\Sigma^0$ dominates
over the one with $\Lambda$ in the high $Q^2$ region.
This is particularly noticeable for $F_2^Y$.

The similarity between the results
for $\gamma^\ast \Lambda \to \Lambda^\ast$
and  $\gamma^\ast \Sigma^0 \to \Lambda^\ast$
can be understood by the expressions
for the form factors given by Eqs.~(\ref{eqF1})-(\ref{eqF2})
and Eqs.~(\ref{eqF1s})-(\ref{eqF2s}).
In both cases there is a dependence
on the overlap integral ${\cal I}_Y$.
As the scalar wave functions have
the same parametrization for the $\Lambda$ and $\Sigma^0$
the difference in the overlap integrals in their respective rest frames
are only due to the masses ($M_\Lambda$ and $M_\Sigma$),
leading to almost the same results for both cases.
Therefore, the main difference in the
form factors comes from the flavor
factors that are multiplied by the overlap integrals.
Although the flavor factors contain the functions
$f_{i+}, f_{i-}$ and $f_{i 0}$ ($i=1,2$)
which are dependent on $Q^2$,
we can make a simple estimate in the
exact $SU(3)$ limit taking $f_{i+} = f_{i-} = f_{i 0}$ ($i=1,2$).
In this limit we have
$F_i^\Sigma (Q^2) = \sqrt{3} F_i^\Lambda(Q^2)$ ($i=1,2$)
consistent with the magnitude shown in Fig.~\ref{figF1F2}.
We note that our results are
different from those in
Ref.~\cite{VanCauteren05} obtained using
a constituent quark model, and also different from those of
the chiral unitary model~\cite{Doring10} which shows
$|F_2^\Sigma| \gg |F_2^\Lambda|$ for $Q^2 \simeq 0$ as seen later.
This relation comes
from the $\Lambda^\ast$ decay widths to $\gamma\, \Lambda$
($\Gamma_{\gamma \Lambda}$) and $\gamma\, \Sigma^0$
($\Gamma_{\gamma \Sigma^0}$), which are predicted to be
$\Gamma_{\gamma \Sigma^0} \gg \Gamma_{\gamma \Lambda}$
(in general $\Gamma_{\gamma Y} \propto |A_{1/2}^Y(0)|^2 \propto |F_2^Y(0)|^2$).
We recall again that the results of
the present valence quark model are valid for $Q^2 \gg 0.2$ GeV$^2$
and the region near $Q^2=0$ is excluded, and thus we cannot
predict the corresponding decay widths reliably.

We discuss next the rate of the falloff
of the form factors, again apart logarithm corrections.
We measure the falloff based on
$F_1^\ast(Q^2) \approx \left(\sfrac{\Lambda_1^2}{\Lambda_1^2+Q^2} \right)^{2}$
and
\mbox{$F_2^\ast(Q^2) \approx \left(\sfrac{\Lambda_2^2}{\Lambda_2^2+ Q^2}
\right)^{3}$} for $Q^2 \simeq 2$ GeV$^2$.
While for \mbox{$\gamma^\ast \Lambda \to \Lambda^\ast$},
we have $\Lambda_1^2= 3.6$ GeV$^2$ and $\Lambda_2^2= 3.6$ GeV$^2$,
for $\gamma^\ast \Sigma^0 \to \Lambda^\ast$
we have $\Lambda_1^2= 3.1$ GeV$^2$ and $\Lambda_2^2= 3.2$ GeV$^2$.
In all cases, we have
slower falloff than that for the $\gamma^\ast N \to N^\ast(1535)$
reaction.
Since in the flavor symmetric limit the falloff should
be same among the octet baryons,
the differences among $N$ and $\Lambda$ (or $\Sigma^0$),
and $N^{*}$ and $\Lambda^{*}$,
are a consequence of a special role of
the strange quark which breaks flavor symmetry.

There are two factors that can cause the slower falloff
of the $\gamma^\ast Y \to \Lambda^\ast$ transition form factors
than the one for the $\gamma^\ast N \to N^\ast(1535)$ reaction.
The first one is the difference in the
quark distributions between the
nucleon-$N^\ast(1535)$ and the $Y-\Lambda^\ast$ systems.
The second one is the difference in the
kinematics between the two systems.
As for the difference in the quark distributions,
$\Lambda$ and $\Sigma^0$ are more
compact systems than that of the nucleon,
because they have one heavier strange quark
in contrast with the nucleon which have only
the light quarks.
Therefore, the $\Lambda$ and $\Sigma^0$ are
characterized by the radial wave functions (\ref{eqPsiR})
with a larger extension in the momentum space,
and consequently the overlap integral becomes larger
than the one for the $\gamma^\ast N \to N^\ast(1535)$ reaction.
However, this is not the main factor,
since the parameters corresponding
to the $\Lambda$ and $\Sigma^0$ wave functions are not different
significantly from those of the nucleon\footnote{
While the $\Lambda$ and $\Sigma^0$
radial wave functions are characterized
by the parameter $\beta_3 \simeq 0.76$~\cite{OctetEMFF},
that of the nucleon system is parameterized by
$\beta_2 \simeq 0.72$~\cite{Nucleon}
(smaller momentum scale).
In both cases the additional range parameter
is $\beta_1 \approx 0.05$~\cite{OctetEMFF,Nucleon}.}.
The second factor is the difference in the
kinematics between the two reactions.
For the radial wave functions given by Eq.~(\ref{eqPsiR}),
it is possible to show that the systems
with the same parametrization are
characterized by the overlap integrals ${\cal I}_Y$,
which are functions of the ratio $\sfrac{|{\bf q}|_Y}{M_Y}$
(see Appendix \ref{apInt}).
Therefore, the falloff of ${\cal I}_Y$
is determined by the factor $\sfrac{|{\bf q}|_Y}{M_Y}$.
The larger the ratio, the falloff is larger.
Comparing the values of $\sfrac{|{\bf q}|_\Lambda}{M_\Lambda}$
with the corresponding ratio $\sfrac{|{\bf q}|}{M}$
for the $\gamma^\ast N \to N^\ast(1535)$ reaction,
the latter has the larger ratio for $Q^2=0$
($\sfrac{|{\bf q}|}{M}=0.48$ compared with
$\sfrac{|{\bf q}|_\Lambda}{M_\Lambda}=0.41$)
and this is true for also for larger values of $Q^2$.
This means the overlap integral has stronger falloff for the
$\gamma^\ast N \to N^\ast(1535)$,
and it is reflected on the faster falloff of the form factors.

To compare the falloff for the reactions
$\gamma^\ast \Lambda \to \Lambda^\ast$
and $\gamma^\ast \Sigma^0 \to \Lambda^\ast$,
we need to analyze more in detail
to explain the difference in
the observed behavior, namely, the falloff for
the reaction involving $\Sigma^0$
is faster (smaller cutoffs) than that for
the reaction involving $\Lambda$.
In this case the ratios $\sfrac{|{\bf q}|_Y}{M_Y}$
are close, $\sfrac{|{\bf q}|_\Lambda}{M_\Lambda}= 0.41$
and $\sfrac{|{\bf q}|_\Sigma}{M_\Sigma}= 0.36$,
for $Q^2=0$.
The important effect now is the
contribution from the flavor factors
in the form factors, given by
Eqs.~(\ref{eqF1})-(\ref{eqF2}) and (\ref{eqF1s})-(\ref{eqF2s}).
The form factors for the reaction involving the $\Lambda$
have dependence on the strange quark form factors $f_{10}$ and $f_{20}$,
and that these functions have slower falloff with $Q^2$
than those with the reaction involving the $\Sigma^0$,
which depend only on the light quark form factors.
Then, the corresponding transition form factors
for the reaction with $\Lambda$ also have slower
falloff than that for the reaction with $\Sigma^0$.

\subsubsection{Results of chiral unitary model}

\begin{figure}[t]
\centerline{
\includegraphics[width=1.7in]{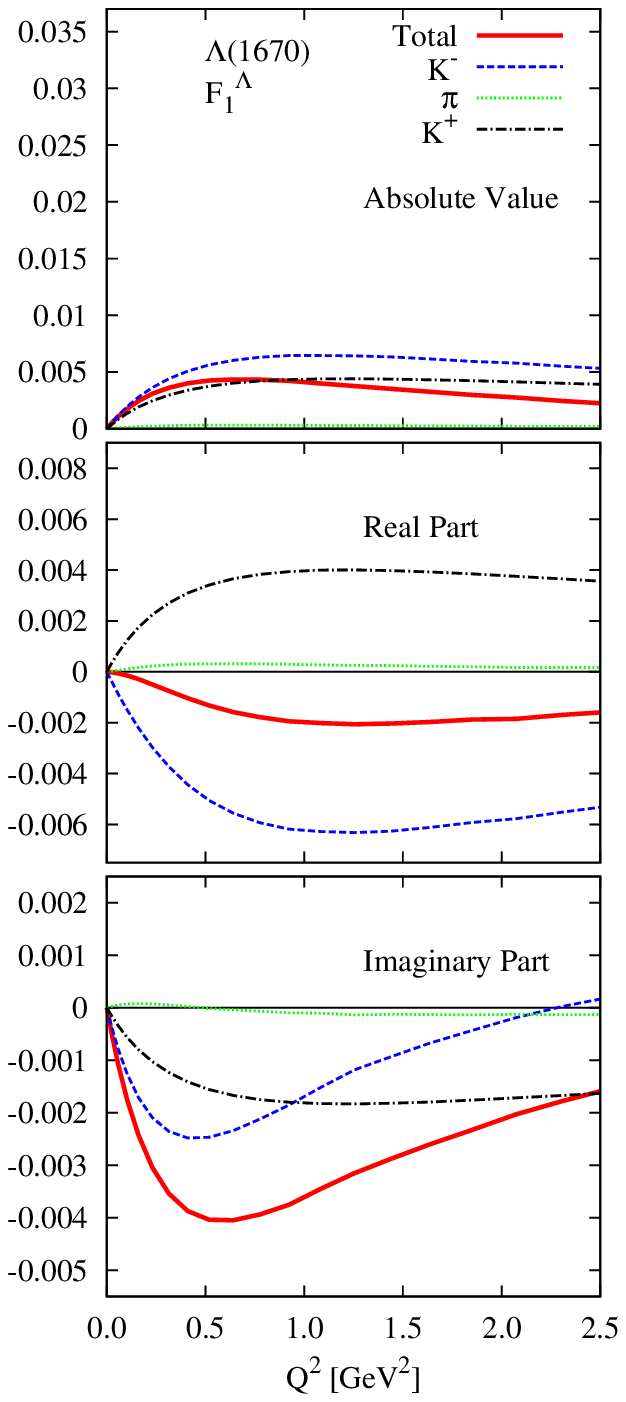}
\includegraphics[width=1.7in]{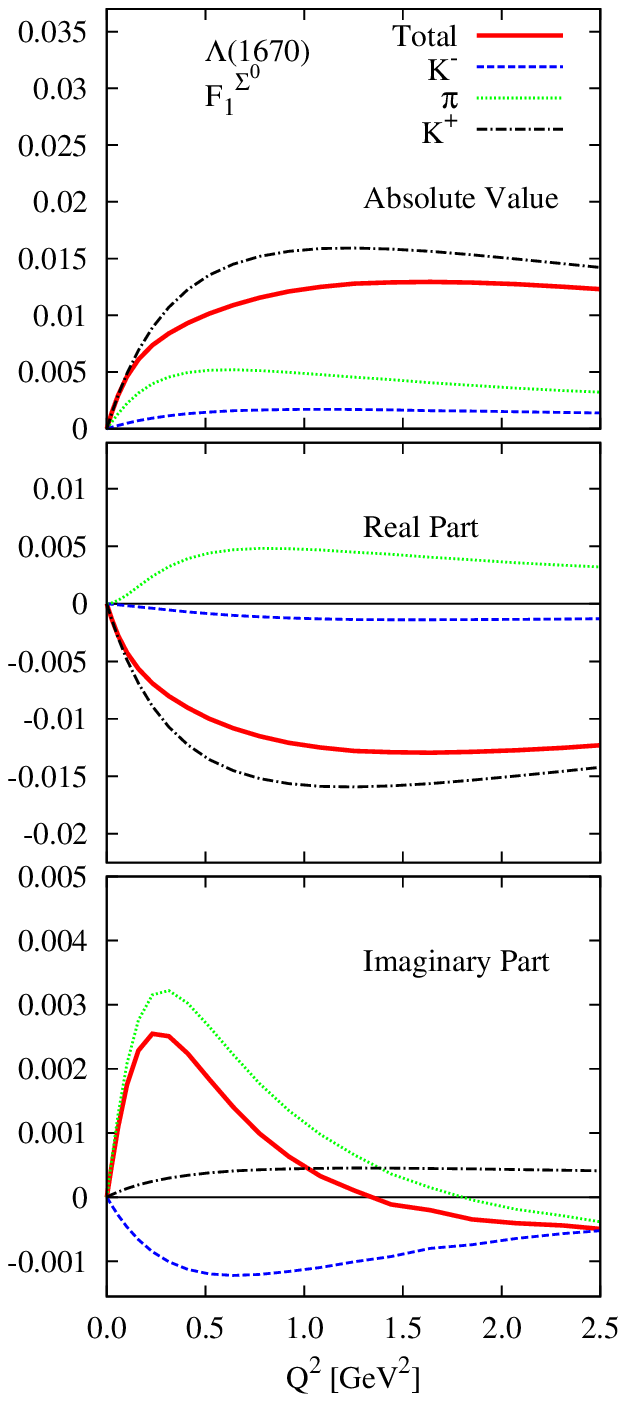}
}
\caption{\footnotesize{(Color online)
Meson cloud contributions
for the Dirac-type form factor $F_{1}^{Y}(Q^{2})$ of
the $\gamma^\ast Y \to \Lambda^\ast$ transition
for $Y=\Lambda$ (left panels) and $Y=\Sigma^0$ (right panels).
The solid line shows the total contribution coming from diagram (a), while
the dashed, dotted and dot-dashed lines denote the $K^{-}$, $\pi$ and $K^{+}$
contributions, respectively.  }}
\label{figF1ChUM}
\end{figure}

\begin{figure}[t]
\centerline{
\includegraphics[width=1.7in]{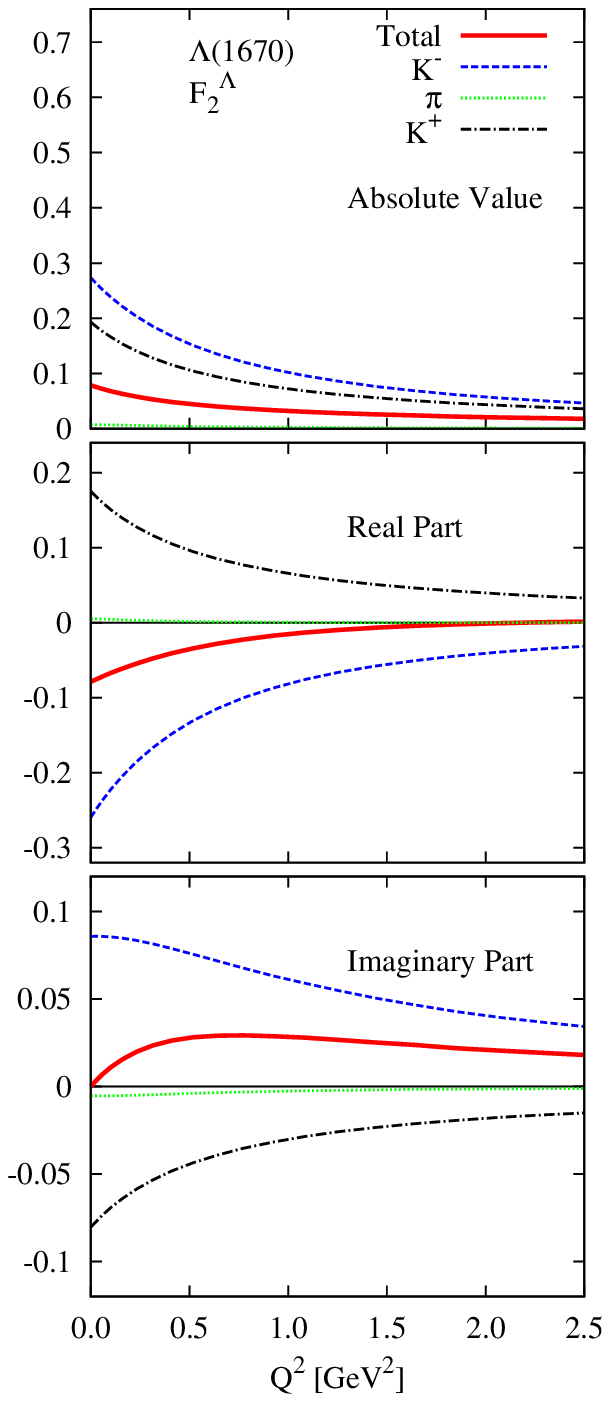}
\includegraphics[width=1.7in]{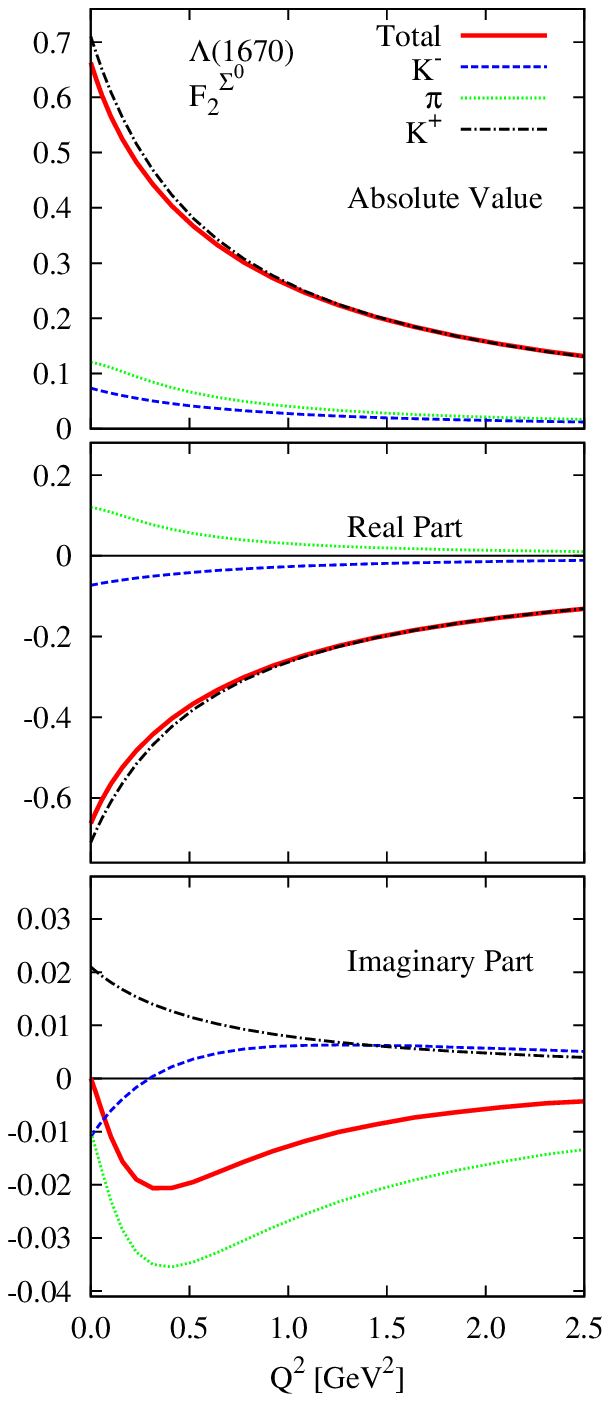}
}
\caption{\footnotesize{(Color online)
Meson cloud contributions
for the Pauli-type form factor $F_{2}^{Y}(Q^{2})$ of
the $\gamma^\ast Y \to \Lambda^\ast$ transition
for $Y=\Lambda$ (left panels) and $Y=\Sigma^0$ (right panels).
The solid line shows the total contribution coming from diagram (a), while
the dashed, dotted and dot-dashed lines denote the $K^{-}$, $\pi$ and $K^{+}$
contributions, respectively.  }}
\label{figF2ChUM}
\end{figure}

Next we show the result of the transition form factors calculated in the chiral unitary
model. In Figs.~\ref{figF1ChUM} and \ref{figF2ChUM}, we show the results of the Dirac- 
and Pauli-type form factors, $F_{1}^{Y}(Q^{2})$ and $F_{2}^{Y}(Q^{2})$,
for the $\gamma^{*} Y \to \Lambda^{*}$ transition, respectively. 
As seen in the figures, in the meson cloud model the $F_{2}^Y$ form factors
for both cases are one order of magnitude larger than the $F_{1}^Y$
form factors. This is very similar tendency as that for the $N(1535)$ case.

In Figs.~\ref{figF1ChUM} and \ref{figF2ChUM} we also show the contribution coming from each meson separately. For the $\Lambda$ transition form factors the pion contribution is very small. This is because only the isoscalar component of the photon current can contribute the $\gamma^\ast \Lambda \to \Lambda^{*}$
transition in the isospin symmetric limit, while the pion with isospin 1 can couple only to the isovector part of the photon current. Thus, with very small isospin breaking effect there is little pion cloud contribution in the
$\gamma^\ast \Lambda \to \Lambda^{*}$ transition.
It is also interesting to mention that for the $\gamma^\ast \Lambda  \to \Lambda^{*}$ transition there is cancellation between $K^{-}$ and $K^{+}$ contributions, while the $\gamma^\ast \Sigma^{0}  \to \Lambda^{*}$ transition is dominated by the $K^{+}$ cloud component.
In this way, for the both form factors $F_{1}^Y$ and $F_{2}^Y$,
the transition from $\Sigma^{0}$ is larger than that from $\Lambda$.
Especially for
the Pauli-type form factor, $F_{2}^{\Sigma}$
is almost seven times larger than $F_{2}^{\Lambda}$. 
This cancellation is also found for the helicity amplitudes
(see Ref.~\cite{Doring10} for the details).

\subsubsection{Comparison of two models}

\begin{figure*}[t]
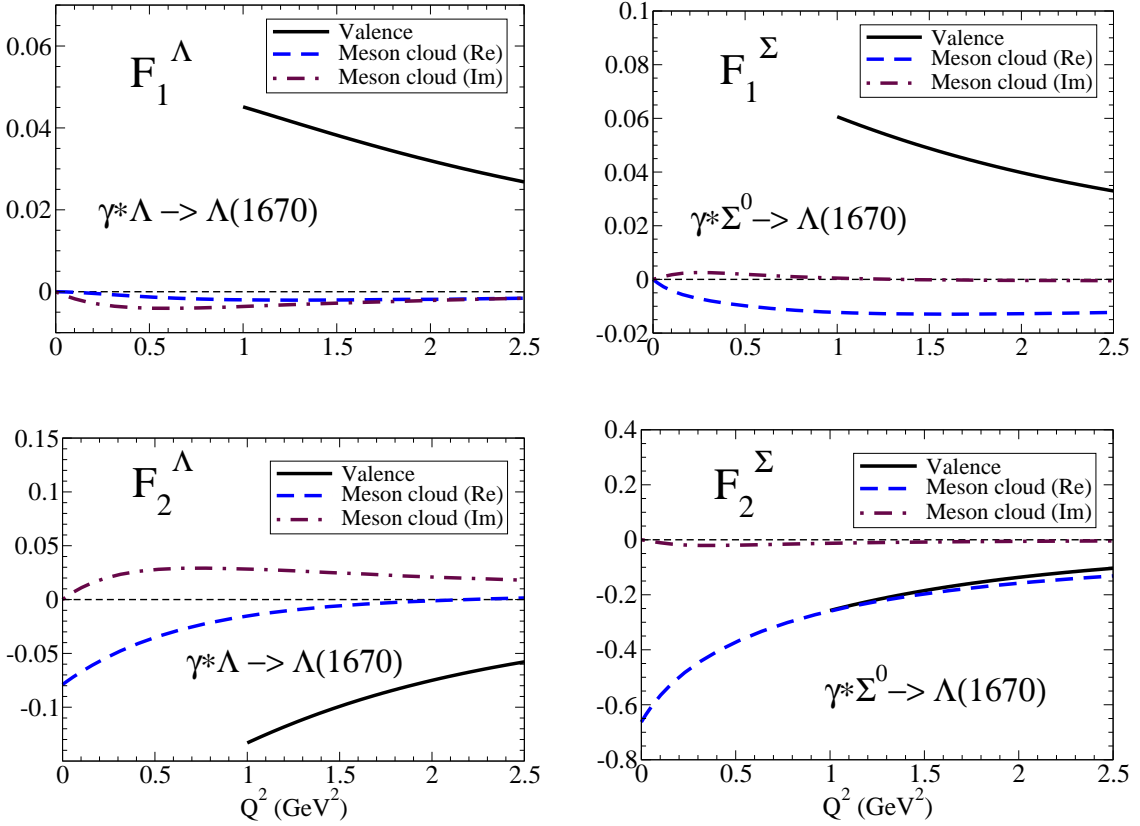

\vspace{.4cm}
\centerline{
\includegraphics[width=2.8in]{F1lamb3.eps}
\hspace{.5cm}
\includegraphics[width=2.8in]{F1sigma1.eps}
}
\vspace{.8cm}
\centerline{
\includegraphics[width=2.8in]{F2lamb3.eps}
\hspace{.5cm}
\includegraphics[width=2.8in]{F2sigma1.eps}
}
\caption{\footnotesize{ (Color online)
Valence quark and meson cloud contributions
for the $\gamma^\ast Y \to \Lambda^\ast$ transition form factors
for $Y=\Lambda$ (left panel), and $Y=\Sigma^0$ (right panel).}}
\label{figLamb1670}
\end{figure*}

Finally, we compare the valence quark contributions with
those from the meson cloud for the reactions
involving the $\Lambda$ and $\Sigma^0$.
The comparison is shown in Fig.~\ref{figLamb1670},
for the reactions involving the $\Lambda$ (left panel)
and the $\Sigma^0$ (right panel).
Note that the chiral unitary model results
have both the real and imaginary parts and
the absolute phases are fixed at $Q^{2}=0$ to give
a real and positive value of $A_{1/2}^{Y}(0)$,
or equivalently real and negative $F_{2}^{Y}(0)$, as
mentioned before.

One can see in Fig.~\ref{figLamb1670}
that the imaginary part is small in general.
Therefore, we will focus only on the real part hereafter.
Another interesting point
is that, while the both form factors for the reaction
involving the $\Lambda$ are dominated by the valence quark
contributions, this is not the case for the reaction involving the $\Sigma^0$.
For the latter case one can see that the valence quark contributions
for $F_1^\Sigma$ are larger than those from the meson cloud
(about 2.5 times near $Q^2=2.5$ GeV$^2$),
although the magnitude is similar for $F_2^\Sigma$.

A point of particular importance about $F_2^{\Sigma}$
is the relative sign between the valence and meson cloud contributions.
Since, as mentioned before, the factor $\eta_{\Lambda \Sigma^0}$ is unknown
at present, we cannot decide if there is
a positive or negative interference between the contributions. 
The results for the form factors involving the $\Sigma^0$ are
determined using $\eta_{\Lambda^\ast} \eta_{\Lambda \Sigma^0}=1$.
If this is the case, there is a combination
of the signs to enhance the total magnitude of $F_2^\Sigma$.
On the other hand, if the sign is opposite,
$\eta_{\Lambda^\ast} \eta_{\Lambda \Sigma^0}=-1$,
one can expect a cancellation between
the valence quark and meson cloud effects,
leading to a result $F_2^\Sigma \approx 0$,
or to a magnitude similar to $F_1^\Sigma$.
An example of a quark model with $F_2^\Sigma >0$
can be found in Refs.~\cite{VanCauteren05,VanCauteren03}.
Note that in case $\eta_{\Lambda^\ast} \eta_{\Lambda \Sigma^0}=-1$,
the reaction $\gamma^\ast \Sigma^0 \to \Lambda^\ast$
has similar properties with
the reaction $\gamma^\ast N \to N^\ast(1535)$,
discussed previously.
This result suggests that the experimental determination of
the sign for $F_2^\Sigma$ is very important to
pin down the relative phase between the $\Lambda$
and $\Sigma^0$ wave functions in the present model
as we explain next.

From the discussions above, we conclude that,
if $\eta_{\Lambda^\ast} \eta_{\Lambda \Sigma^0}=+1$,
it is expected that $F_2^\Sigma$ becomes larger in magnitude.
In the alternative case, $\eta_{\Lambda^\ast} \eta_{\Lambda \Sigma^0}=-1$,
$F_2^\Sigma$ should be smaller in magnitude, 
and comparable with $F_1^\Sigma$. 
Therefore, once the sign $\eta_{\Lambda^\ast}$ is known,
$\eta_{\Lambda \Sigma^0}$ can be inferred from
the result for $F_2^\Sigma$. 
Note also that in our model $\eta_{\Lambda^\ast}$
can be fixed by the results for the reaction
$\gamma^\ast \Lambda \to \Lambda^\ast$, since
valence quark effect dominates that transition
($F_1^\Sigma \propto \eta_{\Lambda^\ast}$).
Then $\gamma^\ast \Lambda \to \Lambda^\ast$
can be used to determine $\eta_{\Lambda \Sigma^0}$,
which fixes also the sign of $\mu_{\Lambda \Sigma^0 }$.

The covariant spectator quark model
can also be used to calculate
the valence quark contributions for the
$\gamma^\ast \Lambda \to \Sigma^0$ form factors
in general and the transition magnetic moment
$\mu_{\Lambda \Sigma^0}$ in particular.
Assuming that the valence quark effect is
the leading contribution
as demonstrated reasonable for the octet baryon system~\cite{OctetEMFF},
one can conclude that $\mu_{\Lambda \Sigma^0 }
\propto -\eta_{\Lambda \Sigma^0}$~\cite{InPreparation},
namely, the sign of $\mu_{\Lambda \Sigma^0 }$ is the opposite
to that of $\eta_{\Lambda \Sigma^0}$.
Thus, once determined the sign of $\eta_{\Lambda \Sigma^0}$
corresponding to the reaction $\gamma^\ast \Sigma^0 \to \Lambda^\ast$,
we can determine the sign of $\mu_{\Lambda \Sigma^0}$.

\begin{figure*}[t]
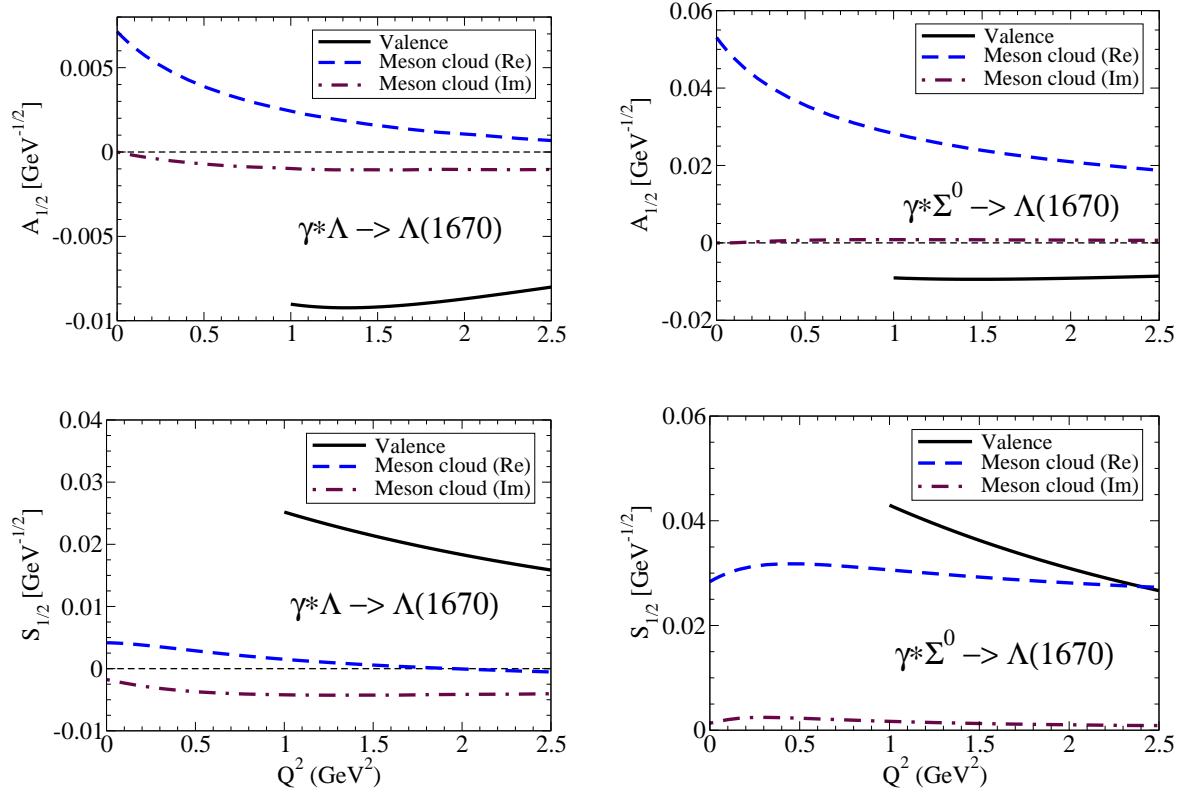

\vspace{.4cm}
\centerline{
\includegraphics[width=2.9in]{A12_L1670.eps}
\hspace{.5cm}
\includegraphics[width=2.9in]{A12_S1670.eps}
}
\vspace{.8cm}
\centerline{
\includegraphics[width=2.9in]{S12_L1670.eps}
\hspace{.5cm}
\includegraphics[width=2.9in]{S12_S1670.eps}
}
\caption{\footnotesize{ (Color online)
Valence quark and meson cloud contributions
for the $\gamma^\ast Y \to \Lambda^\ast$ helicity amplitudes
for $Y=\Lambda$ (left panel) and $Y=\Sigma^0$ (right panel).}}
\label{figHelicities}
\end{figure*}

For completeness, we also present in Fig.~\ref{figHelicities}
the results for the helicity amplitudes $A_{1/2}^Y$ and $S_{1/2}^Y$
converted from $F_{1}^{Y}$ and $F_{2}^{Y}$ by Eqs.~(\ref{eqA12X}) and~(\ref{eqS12X}),
assuming the same phases as the form factors\footnote{The $A_{1/2}^{Y}$ helicity amplitude shown in Ref.~\cite{Doring10} has a different absolute phase from the present work.}.
For the $\gamma^\ast \Lambda \to \Lambda^\ast$ reaction
we can also observe the dominance of the valence quark
contributions over the meson cloud contributions.
As for the $\gamma^\ast \Sigma^0 \to \Lambda^\ast$ reaction,
the more interesting point is the closeness of
the valence and meson cloud contributions
for the $S_{1/2}^\Sigma$ amplitude.
Also in this case we can conclude that
if $\eta_{\Lambda^\ast} \eta_{\Lambda \Sigma^0} = + 1$,
$S_{1/2}^\Sigma$ is enhanced,
while the alternative case, $\eta_{\Lambda^\ast} \eta_{\Lambda \Sigma^0} =-1$,
we expect a substantial reduction of the $S_{1/2}^\Sigma$ amplitude.
The similarity in the behavior for $S_{1/2}^\Sigma$
and $F_2^\Sigma$, as discussed before,
is a consequence of the partial suppression
of the $F_1^\ast$ contribution for the $S_{1/2}^\Sigma$ amplitude,
due to the factor $\sfrac{M_{\Lambda^\ast}-M_Y}{M_{\Lambda^\ast} + M_Y}$.
Another interesting point
in Fig.~\ref{figHelicities}
is the flatness of the valence
quark model result for $A_{1/2}^Y$ as a function of $Q^2$
around the region $Q^2=2$ GeV$^2$.
This is because the region $Q^2=2$ GeV$^2$ is the turning point of
changing the $Q^2$ dependence in the amplitude.
For the larger $Q^2$ region however,
the expected falloff with the $Q^2$ can be observed.

\section{Conclusions}
\label{secConclusions}

In this study we have analyzed the contributions from
the valence quark and meson cloud effects
for the $\gamma^\ast B \to B^\ast$  reactions
with $B=N,\Lambda,\Sigma^0$ and  $B^\ast=N(1535),\Lambda(1670)$.
While the valence quark contributions are estimated
using a constituent quark model~\cite{Nucleon,S11,OctetEMFF},
those of the meson cloud are estimated using
the chiral unitary model~\cite{Jido08,Doring10}.
In the chiral unitary model, the $N(1535)$ has some components other than
meson-baryon dynamics as discussed in Ref.~\cite{Hyodo:2008xr}, but
for the calculation of the transition form factors we take only coupling of
the photon current to the meson component and do not take into account
of photon couplings to genuine quark components.
In this approach the $\Lambda(1670)$ is
almost composed of meson-baryon components~\cite{Doring10}. 
Since the valence and meson cloud
effects are calculated by the different formalisms
we cannot simply combine the both contributions to
obtain the final, total results for the transition form factors.
Nevertheless, the magnitude and signs
of the individual contributions presented here
are sufficient to conclude that it
is possible to have a cancellation from
the two effects, the valence quark and meson cloud effects,
in a consistent, unified approach including the both effects.

For the $\gamma^\ast N \to N^\ast(1535)$ reaction,
we have found difference in signs
for the two contributions for the Pauli-type form factor $F_2^\ast$,
which can be the main reason for the experimental observation,
$F_2^\ast \simeq 0$ for $Q^2 > 2$ GeV$^2$.

As for the reactions $\gamma^\ast Y \to \Lambda(1670)$
($Y=\Lambda,\Sigma^0$), we conclude
that generally the valence quark contributions dominate
for the $Y=\Lambda$ case, but the two
contributions are similar for the reaction with $Y=\Sigma^0$.
A particularly interesting case is the
form factor $F_2^\Sigma$.
Namely, if we assume the same sign
for the $\Lambda$ and $\Sigma$
radial wave function normalization constants
($\eta_{\Lambda \Sigma^0} =1$) and $\eta_{\Lambda^\ast}=1$,
we have an enhancement for $F_2^\Sigma$.
Instead, if we assume $\eta_{\Lambda^\ast} \eta_{\Lambda \Sigma^0} =-1$,
we have a substantial cancellation
between the two effects.
Then, the $F_2^\Sigma$ contribution for the
reaction cross section would be very small.

A consequence of the observation made above
is that the $\gamma^\ast \Sigma^0 \to \Lambda^\ast$
reaction can provide an indirect method to determine
$\eta_{\Lambda \Sigma^0}$,
which can be used to pin down the sign of
$\mu_{\Lambda \Sigma^0}$ consistently within the present approach.
This can be of fundamental importance,
because the sign of the
$\gamma^\ast \Lambda \to \Sigma^0$ transition
form factors, and in particular the
sign of the transition magnetic moment,
$\mu_{\Lambda \Sigma^0 }$, is not determined
experimentally.
Also this sign has not been related consistently
with the other reactions so far.
Although the sign is predicted to be negative within the
unitary symmetry approach~\cite{Coleman}
(the same sign with the neutron magnetic moment),
the consistency with other reaction was not studied
within the approach.

From the discussion made above we conclude that
the theoretical and experimental studies
of the reactions $\gamma^\ast N \to N^\ast(1535)$
and $\gamma^\ast Y \to \Lambda^\ast$, with $Y=\Lambda,\Sigma^0$,
as well as the correlations between them,
are very interesting topics of investigation.
The results from these transition form factors
can be used to estimate the light and
strange quark distributions in the baryons,
as well as to predict other reactions.

\vspace{0.2cm}
\noindent

\begin{acknowledgments}
G.~R.~acknowledges CSSM, the University of Adelaide for making him
possible to visit and stay, and thank A.~W.~Thomas and
H.~Kamano for helpful discussions.
D.~J.~would like to thank M.~Doring and E.~Oset for the collaboration
on the helicity amplitudes in the chiral unitary model.
This work was partially financed by the European Union
(HadronPhysics2 project ``Study of strongly interacting matter'')
and by the Funda\c{c}\~ao para a Ci\^encia e a
Tecnologia, under Grant No.~PTDC/FIS/113940/2009,
``Hadron structure with relativistic models'', and
partially supported by the Grants-in-Aid
for Scientific Research (No.~22740161 and No.~22105507),
and carried out in part under the Yukawa International
Program for Quark- hadron Sciences (YIPQS).
This work was also supported by the University of Adelaide and
the Australian Research Council through grant No. FL0992247 (AWT).
G.~R.\ was supported by the Funda\c{c}\~ao para
a Ci\^encia e a Tecnologia under the Grant
No.~SFRH/BPD/26886/2006.
\end{acknowledgments}

\appendix

\section{Current matrix elements in the covariant spectator quark model}
\label{JY}

The calculation of the matrix elements
for a transition between a $J^P=\sfrac{1}{2}^+$
initial state and a $J^P=\sfrac{1}{2}^-$ final state
follows the same steps as that of Appendix B in Ref.~\cite{S11}
for the $\gamma^\ast N \to N(1535)$ reaction.
Here it is sufficient to note that
in the transition current
involving the $\Psi_{\Lambda^\ast}$ the
terms in $(\varepsilon_\pm \cdot \tilde k)$
vanish in the $k$ integral.
Therefore, only the terms proportional
to the integral
\ba
{\cal I}_{Y} =
- \eta_{\Lambda^\ast}
\int_k N(\varepsilon_0 \cdot \tilde k) \psi_{\Lambda^\ast}\psi_Y,
\ea
survive in the current.
The minus sign is introduced for convenience.

With this simplification
we can derive the following results,
\ba
& &
\hspace{-1.3cm}
\int_k \left[ \,
\overline \Phi_\rho \hat \gamma^\mu \phi_S^0 \right] \!
\psi_{\Lambda^\ast} \psi_Y
=
-{\cal I}_Y
\left\{
\bar u_{\Lambda^\ast}  \hat \gamma^\mu
\gamma_5 u_Y
\right\},
\\
& &
\hspace{-1.3cm}
\int_k
\left[
\overline \Phi_\rho
  \frac{i \sigma^{\mu \nu} q_\nu}{2M}
 \phi_S^0 \right]
\!\psi_{\Lambda^\ast} \psi_Y =
 {\cal I}_Y
\left\{
\bar u_{\Lambda^\ast}
 \frac{i \sigma^{\mu \nu} q_\nu}{2M}
\gamma_5 u_Y \right\}, \\
& &
\hspace{-1.3cm}
\int_k
\left[
\overline \Phi_\lambda  \hat \gamma^\mu \phi_S^1\right]
\!\psi_{\Lambda^\ast}\psi_Y=
\frac{1}{3}  {\cal I}_Y
\left\{
\bar u_{\Lambda^\ast}  \hat \gamma^\mu \gamma_5 u_Y
\right\}, \\
& &
\hspace{-1.3cm}
\int_k
\left[
\overline \Phi_\lambda
  \frac{i \sigma^{\mu \nu} q_\nu}{2M}  \phi_S^1\right] \!
\psi_{\Lambda^\ast}\psi_Y=
-\frac{1}{3} {\cal I}_Y
\left\{
\bar u_{\Lambda^\ast}  \frac{i \sigma^{\mu \nu} q_\nu}{2M} \gamma_5 u_Y
\right\}.
\ea

Inserting these results into the
expression of the current, we obtain
\ba
J_Y^\mu &=& + e\, \frac{1}{2}(3 j_1^A + j_1^S)
{\cal I}_Y \,
\hat \gamma^\mu  \gamma_5
\nonumber \\
& &- e\, \frac{1}{2}(3 j_2^A - j_2^S)
{\cal I}_Y \,
\frac{i \sigma^{\mu \nu} q_\nu}{2M}  \gamma_5.
\ea

\section{Overlap integral}
\label{apInt}

Consider the overlap integral in the
final state ($\Lambda^\ast$) rest frame,
given by Eq.~(\ref{eqInt2}),
with the radial wave functions of Eq.~(\ref{eqPsiR}),
\ba
{\cal I}_Y(Q^2)&=&
\eta_{\Lambda^\ast}
\frac{
{\cal N}_{\Lambda} {\cal N}_Y}{m_D^2}
\int_k \frac{k_z}{|{\bf k}|}
\left\{
\frac{1}{(\beta_1 + \chi_{\Lambda^\ast})
(\beta_3 + \chi_{\Lambda^\ast})} \right. \nonumber \\
& &
\left.
\times
\frac{1}{(\beta_1 + \chi_Y)  (\beta_3 + \chi_Y )}\right\}.
\label{eqInt3}
\ea
In the above equation $\chi_B$ is determined
by Eq.~(\ref{eqCHI})
for the momenta defined by
Eq.~(\ref{eqLRF}). Therefore,
\ba
\chi_{\Lambda^\ast} 
= 2 \left( \frac{E_D}{m_D} -1\right),
\label{eqCHI1}
\ea
and
\ba
\chi_{Y}
&=& 2 \left( \frac{E_Y}{ M_Y } \frac{E_D}{m_D}+
\frac{|{\bf q}|_Y}{M_Y} k_z -1 \right),
\label{eqCHI2}
\ea
where
\ba
\frac{E_Y}{M_Y}=  \sqrt{ 1 + \frac{ |{\bf q}_Y|^2 }{ M_Y^2 }}.
\ea
From Eq.~(\ref{eqCHI1}) we can see that $\chi_{\Lambda^\ast}$
has no dependence on $Q^2$ (or $|{\bf q}|_Y$),
and that from Eq.~(\ref{eqCHI2})
$\chi_Y$ is a function of the ratio
$\sfrac{|{\bf q}|_Y}{M_Y}$.
Therefore, we can write
\ba
{\cal I}_Y(Q^2) = {\cal I}_Y\left(  \frac{|{\bf q}|_Y}{M_Y}
\right).
\ea
Furthermore, since $\chi_Y$ increases when
$\sfrac{|{\bf q}|_Y}{M_Y}$ increases,
we can conclude that the absolute value of the integrand function
in Eq.~(\ref{eqInt3}) decreases with $\sfrac{|{\bf q}|_Y}{M_Y}$
for a given ${\bf k}$,
and therefore $|{\cal I}_Y|$ decreases
when the ratio $\sfrac{|{\bf q}|_Y}{M_Y}$ increases.
These results show that,
when we have two reactions described
by the same radial wave function
(the same values for the parameters $\beta_1$ and $\beta_3$),
the reaction with larger ratio $\sfrac{|{\bf q}|_Y}{M_Y}$
for a given $Q^2$, gets the smaller value for $|{\cal I}_Y| $.

A simple consequence of the above result
applies for the nucleon-$N^\ast(1535)$
and $\Lambda-\Lambda^\ast$ transition form factors,
when the wave functions are parameterized exactly the same,
$\sfrac{|{\bf q}|}{M}$ for the nucleon case
is larger than $\sfrac{|{\bf q}|_\Lambda}{M_\Lambda}$
for the $\Lambda$ case, and we have,
\ba
|{\cal I}_\Lambda (Q^2)| > |{\cal I}_N (Q^2)|.
\ea
This relation also explains the faster falloff
of the $\gamma^\ast N \to N^\ast(1535)$ transition
form factors than that of
the $\gamma^\ast \Lambda \to \Lambda^\ast$
transition form factors.

\end{document}